\documentclass[12pt]{iopart}
\usepackage{graphicx}
\usepackage{epsfig}
\usepackage{color}
\usepackage{mathptmx}                
\usepackage{dcolumn}                 
\usepackage{bm}                      
\usepackage{sistyle} 			
\usepackage{eurosym} 			
\usepackage{soul}
\usepackage{latexsym}
\usepackage{amssymb}
\usepackage{multirow} 
\usepackage{mathrsfs} 
\usepackage{booktabs}
\usepackage{longtable}
\usepackage{rotating}
\usepackage{epstopdf}
\usepackage{url}
\usepackage{mcite}
\usepackage{inputenc}
\usepackage{mathpazo}

\def\beqa{\begin{equation}}
\def\eeqa{\end{equation}}
\def\beqaa{\begin{eqnarray}}
\def\eeqaa{\end{eqnarray}}

\begin{document}

\title[Parametrization of the surface energy...]
{Parametrization of the surface energy in the ETF approximation}
\author{U J Furtado$^{1,2}$ and F Gulminelli$^{1}$}
\address{$^1$ Normandie Univ., ENSICAEN, UNICAEN, CNRS/IN2P3, LPC Caen, F-14000 Caen, France.}
\address{$^2$ Depto de F\'{\i}sica - CFM - Universidade Federal de Santa Catarina,  Florian\'opolis - SC - CP. 476 - CEP 88040-900 - Brazil.}
\ead{ujfurtado@gmail.com}

\begin{abstract}
We perform extended Thomas-Fermi calculations in the Wigner-Seitz cell below and above neutron drip
with realistic  functionals. The resulting energy density is decomposed as a sum of bulk terms and a surface term, 
and a compressible liquid drop analytical formula is used to fit the surface tension.  
The effect of curvature terms and neutron skin is studied in detail. 
A very good reproduction of the microscopic data is obtained using an expression that depends only on the mass and charge of the cluster, showing that the in-medium modifications of the nuclear energy in the presence of an external neutron gas can be effectively accounted for in the isospin dependence of the surface tension. 
In this first application, aimed at establishing the fitting protocol, we concentrate on the Sly4 energy functional, but the study can be easily generalized to different functionals, and the resulting parametrizations can be used for direct applications in pasta calculations in neutron star crusts and supernova matter.
\end{abstract}


\section{Introduction}\label{intro}

A reliable quantitative estimate of the surface tension of atomic nuclei is a problem as old as nuclear physics, and it is not yet available in spite of the fact that the theoretical tools needed to address the question have been developed since the early eighties \cite{Brack1985}. They mainly consist
of compressible liquid drop models \cite{Myers1985} (CLDM) at different levels of sophistication, 
including phenomenological parameters that are optimized on nuclear data \cite{Audi2017} or 
microscopic calculations in the  Hartree-Fock-Bogoliubov (HFB) \cite{Goriely2013}, Hartree-Fock (HF), extended Thomas-Fermi (ETF), or Thomas-Fermi (TF) \cite{Centelles1990,onsi2008} approaches. 
In particular, many authors have chosen to optimize  the surface parameters using theoretical calculations rather than experimental data \cite{Aboussir1995,Danielewicz2009,Lee2011,Nikolov2011}.  This is because this strategy allows a better extrapolation to regions not covered by experimental measurements, keeping  at the same time an excellent level of reproduction of measured masses, since the microscopic models are themselves continuously updated and optimized on experimental data and improved ab-initio calculations. 
This is particularly true for astrophysical applications, where the nuclei of interest systematically lay close, or even above, the neutron dripline and can only be accessed through theoretical calculations. This same strategy will be employed in the present paper.

Focusing on recent applications, it was shown that the surface tension impacts the behaviour of unified equations of state (EoS) \cite{Fortin2016} for the description of neutron stars, mergers and supernova matter \cite{Oertel2017,Burgio2018}.
Not all such works require the explicit knowledge of the surface tension. Indeed, if one is solely interested in zero temperature catalyzed neutron star matter, or finite temperature matter in the single nucleus approximation, it is possible to build a unified EoS in a fully microscopic approach, without introducing explicit surfaces and the associated surface tension. This is the case of the first TF \cite{Shen1998,maruyama2006,maruyama2005,Avancini2008} and HF calculations \cite{grygorov2010}, as well as for the most recent works with up-to-date functionals \cite{BCPM2015,Pearson2020}.

However, for direct use in astrophysical applications, the EoS must be provided in an analytical \cite{Potekhin2013} 
or tabulated \cite{compose} form. For this reason, simpler approaches using CLDM  are  mostly used for astrophysical purposes, 
as it is done, for example, in the case of the Sly4 based Douchin-Haensel (DH) EoS \cite{Douchin2001} for the neutron star crust, and the Lattimer-Swesty (LS) EoS \cite{LS1991} for sub-saturation supernova matter, also based on interactions of the Skyrme family. Moreover, the use of a parametrized surface tension becomes compulsory if one wants to make systematic calculations with a large number of functionals to address the model dependence of the results \cite{Fortin2016,Lim2018,*Lim2019a,*Lim2019b,Chatterjee2017,Carreau2019a,*Carreau2019b}, and if one wants to describe the full cluster distribution  associated with finite temperature matter including nuclei at and beyond the dripline \cite{Grams2018,Raduta2019,Barros2020}.

To perform systematic studies with several different functionals, the most flexible solution would be an analytical surface tension expressed only in terms of the functional parameters. For this purpose, analytical solutions of the ETF variational problem with parametrized density profiles were developed in the past  \cite{Krivine1983,treiner1986,Aymard2016a}. However, strong approximations are needed to treat asymmetric nuclei, which prevent the use of such analytical models in very neutron rich matter \cite{Aymard2016b,Chatterjee2017}. A pragmatic solution was adopted in \cite{Carreau2019a,*Carreau2019b}: given a functional, bulk properties together with the neutron star core-crust transition density are extracted from infinite nuclear matter calculations, and the (functional dependent) surface parameters are then extracted via a simultaneous fit of measured masses and the transition density.

In this paper, we aim at extracting the surface properties of nuclei present in neutron star matter
directly from functional calculations on finite nuclei. We perform ETF calculations in a Wigner-Seitz (WS) cell of variable size, in order to consider nuclei below and above the neutron dripline, and to explore different values for the density of the dripped neutron gas. This allows us to consider situations that do not correspond to equilibrium configurations in the single nucleus approximation, but occur in the more general multi-component plasma that is expected in finite temperature neutron star matter.   Following previous works \cite{onsi2008,papakonstantinou2013,Aymard2014}, we consider  Fermi-Dirac (FD) profiles for the proton and neutron densities  
inside the cell, and determine the parameters in a variational way, using a meta-modelling approach \cite{margueron2018} for the energy functional; this approach makes it possible to reproduce with high accuracy a very large set of non-relativistic as well as relativistic popular functionals, and interpolate them for advanced statistical studies. To test the method, the parameters corresponding to the popular Sly4 functional are employed in this first study.

We then perform a fit of the optimal energy using a flexible CLDM approach \cite{newton2013}, where the total WS energy is decomposed as a sum of bulk terms (which only depend on the parameters of the functional) and an interface term that, for a given functional, solely depends on the particle numbers of the nucleus. We will show that this simple expression is as accurate as a more complex one that explicitly accounts for different proton and neutron radii, finite size effects on the bulk terms, the presence of particles on the skin, and the density of the outside nucleon gas.  

The plan of the paper is as follows. An outline of the energy functional  and the ETF formalism 
with parametrized density profiles are first presented in section \ref{energy}.
Then the variational method is discussed in section \ref{minimum}, and the results of the minimization with and without a neutralizing electron background, are provided in section \ref{sec ETF results}. We introduce the different parametrizations of the surface energy in section \ref{sec surf formalism}, and give the results of the corresponding fitting protocols in section \ref{sec surf results}. Lastly, the
conclusions are finally drawn in section \ref{sec conclusions}.
Some details are shown in the Appendices.

\section{ETF Formalism}\label{formalism}

\subsection{Energetics in the Wigner-Seitz cell}\label{energy}

In this section, the ETF energy functional used for this study is presented.  
The well-known appealing property of the ETF approximation is that the
non-local terms in the energy density functional are replaced by local gradients, meaning that   
the functional depends solely on the local particle densities. Therefore, the
energy of any arbitrary nuclear configuration can be calculated if the neutron and proton
density profiles $\rho_n(\vec{r})$ and $\rho_p(\vec{r})$ are given.

At second order in the semi-classical non-relativistic $\hbar$ expansion  
\cite{bohigas1976,Brack1985,treiner1986} the energy density reads:

\beqaa 
{\mathcal H}_\mathrm{ETF} [\rho_n,\rho_p]=  
 \sum_{q=n,p} \frac{\hbar^2}{2 m_q^*} \tau_{q} +v + {\mathcal H}_\mathrm{Coul} +
{\mathcal H}_\mathrm{fin}+{\mathcal H}_\mathrm{SO}.
\eeqaa 

Here, $\tau_q$ includes both the local and non-local part of the kinetic energy density for particle type $q=n,p$: 

\beqaa 
\tau_{q} = \tau_{0q}+ \tau_{2q}^l + \tau_{2q}^{nl}; \\
\tau_{0q} = \frac{3}{5} (3 \pi^2)^{2/3} \rho_q^{5/3}; \\
\tau_{2q}^l = \frac{1}{36} \frac{(\nabla \rho_q)^2}{\rho_q} + \frac{1}{3} \nabla^2\rho_q; \\
\tau_{2q}^{nl} = \frac{1}{6} \frac{\nabla \rho_q \nabla f_q}{f_q} + \frac{1}{6} \rho_q \frac{\nabla^2 f_q}{f_q}
-\frac{1}{12} \rho_q \left( \frac{\nabla f_q}{f_q} \right)^2; \\
f_q =\frac{m}{m_q^*}.
\eeqaa 

The density dependent Landau effective masses $m^*_q$   are given in terms of the bare nucleon mass $m$ as:

\begin{equation}
\frac{m}{m^*_q(\rho_n,\rho_p)}=1+\left ( \kappa_\mathrm{sat}+\tau_3\kappa_\mathrm{sym}\delta\right ) \frac{\rho}{\rho_\mathrm{sat}} ,
\end{equation}
 where $\rho=\rho_n+\rho_p$, $\delta=(\rho_n-\rho_p)/\rho$,  $\rho_\mathrm{sat}$ is the saturation density of symmetric nuclear matter, and $\tau_3=1(-1)$ for neutrons (protons).

Concerning the local part of the nuclear potential energy density $v$, we replace the usual local potential Skyrme term:
\begin{equation}
v^{\mathrm{sky}}(\rho_n,\rho_p)=\rho^2 \left ( C_0+D_0\delta^2\right ) + \rho^{\gamma+2}\left ( C_3 + D_3 \delta^2 \right ) , \label{eq:skyrme}
\end{equation}
with a Taylor expansion around saturation, 

\beqaa  \label{pot}
&v(\rho_n,\rho_p)= \rho \sum_{\alpha=0}^N \frac{1}{\alpha !} (v_\alpha^\mathrm{is} + v_\alpha^\mathrm{iv} \delta^2) x^\alpha u_\alpha^N (x),
\eeqaa 
complemented by a low density correction that ensures the correct behaviour at zero density \cite{margueron2018}:
\beqaa 
&u_\alpha^N (x)= 1-(-3 x)^{N+1-\alpha} \exp(-b\rho / \rho_\mathrm{sat}),
\eeqaa 
with $x=(\rho-\rho_\mathrm{sat})/3\rho_\mathrm{sat}$ and $N$ the order of the Taylor development ($N=4$ in this paper).
In this paper we will concentrate on the Sly4 functional \cite{Chabanat1998}, and the $v_\alpha^\mathrm{is}$'s and $v_\alpha^\mathrm{iv}$'s are 
fixed in order to reproduce equation (\ref{eq:skyrme}) with the $C_0,C_3,D_0,D_3,\gamma$ parameters corresponding to Sly4.
The choice of using equation (\ref{pot}) instead of equation (\ref{eq:skyrme}) has no impact on the results presented in this paper, and it is made to open the possibility to make systematic calculations with a large set of functionals in future works.

The Coulomb term ${\mathcal H}_\mathrm{Coul}={\mathcal H}_{\mathrm{c}}+{\mathcal H}_{\mathrm{exc}}$ contains a direct and an exchange term. This latter is evaluated in the Slater approximation \cite{kohn1965,onsi2008}:

\begin{equation}
{\mathcal H}_{\mathrm{exc}} [\rho_e,\rho_p]= -\frac{3e^2}{16 \pi} \left(\frac{3}{\pi} \right)^{1/3}
(\rho_p^{4/3}(r) + \rho_e^{4/3}). \label{eq:exchange}
\end{equation}
 
The direct term is explicitly worked out in spherical symmetry.
For nuclei in the vacuum, that is, in the absence of the electron background, the electromagnetic potential is zero at infinity. The energy density reads:

\beqa
{\mathcal H}_{\mathrm{c}}^{\rho_e=0}[\rho_p]= \frac{e^2}{2} \rho_p(r)
\left[ \int_0^r \rho_p(r') \frac{{r'}^2}{r} dr' + \int_r^\infty \rho_p(r') r' dr'\right]. \label{eq:couldens}
\eeqa

In stellar matter in the WS approximation, the nuclei are organized in a periodic crystal lattice embedded in a uniform electron background gas of density $\rho_e$. 
Charge neutrality is realized in each cell and symmetry arguments impose that the electromagnetic potential be calculated with respect to the centre of the cell. By doing this, one correctly recovers the lattice energy (see \cite{BBP} and \ref{appC}).
We have \cite{onsi2008}:

\beqa
{\mathcal H}_{\mathrm{c}}^{\rho_e}[\rho_p]= \frac{e^2}{2} (\rho_p(r)-\rho_e)
\left[ \int_0^r \rho_p(r') \left( \frac{{r'}^2}{r}-r' \right) dr' + \rho_e \frac{r^2}{6}\right].
\eeqa

The most important non-local interaction terms comprise a surface gradient term and a spin-orbit term.
The surface term  is given by:
\beqa
{\mathcal H}_{\mathrm{fin}}[\rho_n,\rho_p]=
 C_{\mathrm{fin}}\left(\nabla \rho\right)^2 + D_{\mathrm{fin}} \left (\nabla (\rho\delta)\right )^2;
\eeqa
and  the spin-orbit term reads:
\beqaa 
{\mathcal H}_{\mathrm{SO}}[\rho_n,\rho_p]=& -\frac{m}{\hbar^2}W_0^2 \frac{\rho_n}{f_n}\left( (\nabla \rho_n)^2
+ \frac{(\nabla \rho_p)^2}{4} + \nabla \rho_n \nabla \rho_p \right) \nonumber \\
&-\frac{m}{\hbar^2}W_0^2 \frac{\rho_p}{f_p}\left( (\nabla \rho_p)^2
+ \frac{(\nabla \rho_n)^2}{4} + \nabla \rho_n \nabla \rho_p \right).
\eeqaa 
These terms lead to three extra parameters in non-relativistic energy functionals, while they naturally arise in the energy density from the Lagrangian calculation in the case of relativistic mean field theory.

The total energy of a spherical nucleus or cell of radius $R_\mathrm{WS}$ and volume $V_\mathrm{WS}=(4/3)\pi R_\mathrm{WS}^3$,  which is a functional of the densities, is:
\begin{equation}\label{eq total energy}
E_\mathrm{ETF}[\rho_n,\rho_p]= 4 \pi \int_V r^2  {\mathcal H}_\mathrm{ETF}   dr ,
\end{equation}
where the radial integral is extended to the whole space for nuclei in the vacuum, and it is limited  to the cell radius
for nuclei in stellar matter. In this case, the electron density $\rho_e$ is determined by the charge neutrality condition in the cell.

The minimum of this expression for a fixed number of protons and neutrons corresponds to the most stable
configuration.

\subsection{Variational equations}\label{minimum}

The most general way to approach the problem \cite{Centelles1998,Warda2009,Avancini2010,De2012} 
is  to perform a functional derivative in equation (\ref{eq total energy}) and discretize the space  in order to find numerically the values of the local densities, together with the conditions 
that the total baryonic density and the total isospin asymmetry are fixed,
or equivalently that the particle numbers are fixed:

\begin{equation}\label{N fixed}
N_q= 4 \pi \int_V r^2 \rho_q(r) dr.
\end{equation}

Even in the simplified approximation of spherical symmetry, the resulting equations can only be solved numerically.
A simpler approach that has been often employed in the literature \cite{treiner1986,onsi2008,Pearson2012,papakonstantinou2013,Aymard2014,Raduta2018}, and which has shown to give satisfactory results in spherical symmetry, is to write the densities as parametrized profiles
and to perform the minimization with respect to the parameters. We use Fermi-Dirac (FD) profiles:

\begin{equation}\label{densities}
\rho_q(r)= \rho_{bq} + \rho_{Fq}(r), \qquad \rho_{Fq}(r)= \frac{\rho_{cq}}{1+ \exp[(r-R_q)/a_q]}. 
\end{equation}

The eight parameters to be fixed by the energy variation are: the background densities $\rho_{bq}$, which are zero for nuclei in the vacuum; the bulk density parameters $\rho_{cq}$; the diffusivities $a_q$;  the nuclear radii $R_q$.
Particle number conservation allows to determine two out of the total eight parameters. Using FD profiles,  a precise analytical estimate of the integral can be done  at the limit $a_q/R_q \ll 1$ \cite{hasse_book_1988,Aymard2016a}:

\beqaa \label{N fixed 3}
N_{q}- V_\mathrm{WS}\rho_{bq} &= 4 \pi \int_V r^2 (\rho_q(r)-\rho_{bq}) dr \\
 &= \frac{4\pi}{3}\rho_{cq} R_q^3 \left(1+ \pi^2 \frac{a_q^2}{R_q^2} +\Or \left( \left(\frac{a_q}{R_q} \right)^4 \right)  \right), \nonumber
\eeqaa 
leading to:
\beqaa \label{N fixed 2}
R_q= &\left(\frac{3}{4\pi\rho_{cq}}\left (N_{q}-V_\mathrm{WS}\rho_{bq}\right ) \right)^{1/3} \nonumber \\
& \cdot \left[ 1-\frac{\pi^2 a_q^2}{3} \left(\frac{4\pi \rho_{cq}}{3(N_{q}-V_\mathrm{WS}\rho_{bq})} \right)^{2/3}
+\Or \left( \left(\frac{a_q}{R_q} \right)^6 \right) \right].
\eeqaa

\subsubsection{Nuclei in the vacuum}\label{vacuum}

For nuclei in the vacuum, the background densities are zero, and four parameters
remain to be found. We choose:
$a_n,~a_p,~\rho_{cn}$ and $\rho_{cp}$. Since the total number of particles $A=N_n+N_p$ is fixed, the minimization of the energy per particle $E/A$ is equivalent to the minimization of the total energy, and the four equations necessary to have a
unique determination of the parameters are:

\begin{equation}
\frac{\partial E_\mathrm{ETF}}{\partial z_{qi}} = 0; \qquad z_{qi}=a_n,~a_p,~\rho_{cn} ~ \mathrm{and} ~ \rho_{cp}.
\end{equation}

These are simple partial derivatives that commute with the integral in $r$. Thus:

\beqaa 
0=&\int_0^\infty r^2 \left[ \frac{\partial}{\partial z_{qi}} \left(\sum_{q=n,p} \frac{\hbar^2}{2 m_q^*} \tau_{q} \right) \right. 
+ \frac{\partial v}{\partial z_{qi}}   \nonumber \\
 &\left. + C_{\mathrm{fin}} \frac{\partial}{\partial z_{qi}} (\nabla \rho)^2
+ D_{\mathrm{fin}} \frac{\partial}{\partial z_{qi}} (\nabla (\rho\delta))^2
+ \frac{\partial}{\partial z_{qi}} ({\mathcal H}_\mathrm{Coul}+{\mathcal H}_{\mathrm{SO}})
\right ] dr.
\label{eq zero}
\eeqaa 

Detailed expressions are given in \ref{appA}.

\subsubsection{Nuclei in a gas}\label{sec gas}

 For extreme proton-neutron ratios beyond the associated driplines, the lowest energy configuration corresponds to the presence of nucleons in the continuum. If the matter is bound by an external field, these nucleons fill the space outside the nucleus and  the optimal profile in the ETF approximation is  modified. The profile cannot be described any more by a FD function, and a good approximation consists in adding a constant background density term (see equation (\ref{densities})).
This situation is not realized in the laboratory, and it can only occur in stellar matter. The latter is electrically neutral, and electroneutrality is ensured by an electron gas that can be considered as homogeneous \cite{maruyama2005}. Electrons are therefore 
 included as a uniform relativistic Fermi gas. The energy density is given by:

\beqaa 
\varepsilon_e(\rho_e)=&
\frac{1}{\pi^2} \left[\frac{k_{Fe}(k_{Fe}^2+m_e^2)^{3/2}}{4} -
\frac{m_e^2 k_{Fe}\sqrt{k_{Fe}^2 + m_e^2}}{8} \right. \nonumber \\
&-\left. \frac{m_e^4}{8}
\mathrm{ln}\left(\frac{k_{Fe}+\sqrt{k_{Fe}^2 + m_e^2}}{m_e} \right) \right],
\eeqaa 
where $m_e$ is the electron mass and
$k_{Fe}=(3\pi^2 \rho_e)^{1/3}$ is the Fermi momentum of the electrons.
The total electron energy is simply $E_e=\varepsilon_e V_\mathrm{WS}$. 

Because of the dripped component, equation (\ref{N fixed 2}) cannot be used any more to reduce the number of parameters in the variational calculation, unless the $\rho_{bq}$ are added as extra variational parameters.  
For these calculations, we impose a  total baryonic density $\rho$ and a total proton fraction $Y_p$.
Three extra variational variables must be introduced, and we choose $\rho_{bn}$, $\rho_{bp}$ and $R_\mathrm{WS}$. The energy is now a function of seven parameters:
$a_n,~a_p,~\rho_{cn},~\rho_{cp},~\rho_{bn},~\rho_{bp}$ and $R_\mathrm{WS}$.

Since the total baryon number is not fixed with the chosen external constraints, 
the seven equations
necessary to have a unique determination of the parameters are:

\begin{equation}
A\frac{\partial }{\partial z_{qi}} \left(\frac{E_\mathrm{ETF}}{A} \right)= 0, \label{eq:variational}
\end{equation}
where $z_{qi}=a_n,~a_p,~\rho_{cn},~\rho_{cp},~\rho_{bn},~\rho_{bp}$ and $R_\mathrm{WS}.$

These equations are worked out explicitly in \ref{appB}.

\subsection{ETF results} \label{sec ETF results}

\subsubsection{Nuclei in the vacuum}

\begin{figure}
\begin{center}
\includegraphics[scale=1.]{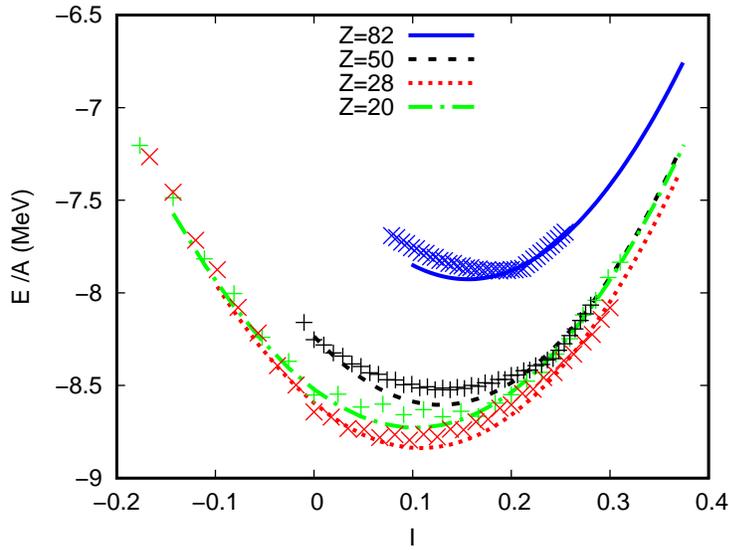}
\end{center}
\caption{Energy per particle for nuclei in the vacuum, from dripline to dripline, for four different isotopic chains corresponding to $Z=82,50,28,20$, as a function of the isospin asymmetry $I=(N-Z)/A$. Lines: ETF results using the Sly4 functional.   
Symbols: experimental data from \cite{Audi2017}.   
}
\label{aenerC}
\end{figure}

\begin{figure}
\begin{center}
\includegraphics[scale=1.]{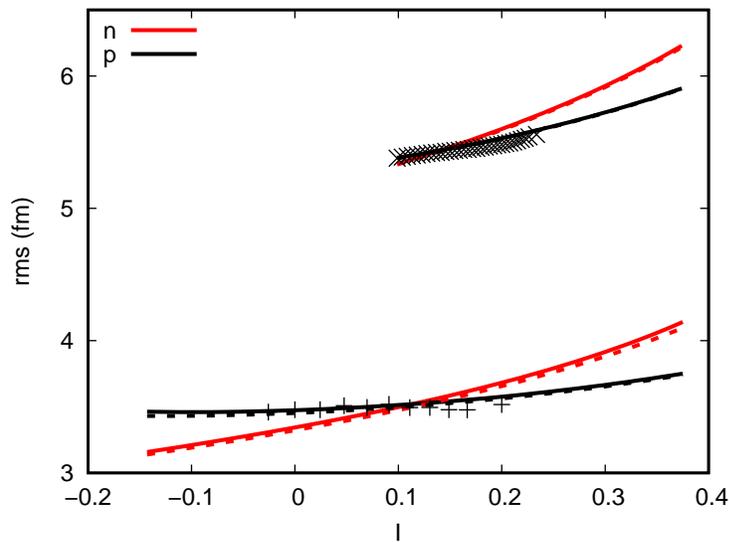}
\end{center}
\caption{Root mean square radii for protons and neutrons for nuclei in the vacuum, as a function of the isospin asymmetry. 
 Dashed lines are for equation  (\ref{eq rms int}) and
solid lines are for equation  (\ref{eq rms exp}). Top lines are for $Z=82$ and bottom lines are for
$Z=20$. Points are experimental data for charge radii from \cite{Marinova2013}.  
}
\label{ArmsC}
\end{figure}

In order to have a general overview of the performance of the model for mass predictions along the nuclear chart, figure  \ref{aenerC} displays the energy per baryon as a function of the isospin asymmetry $I=(N-Z)/A$ along four different isotopic chains. The theoretical ETF calculations are compared to experimental measurements where available. 
 Because of the different approximations employed (spherical symmetry, semi-classical expansion), the model cannot be used for precise applications in nuclear structure, but the global predictive power is comparable to the one of full HF calculations in spherical symmetry \cite{Chatterjee2017}, even if the latter contain shell effects that are neglected here. Concerning the astrophysical applications we are interested in, such precision is certainly insufficient to correctly predict the composition of the outer crust of catalysed neutron stars \cite{Goriely2013,Potekhin2013,Pearson2020}, but we believe it constitutes a sufficient starting point for an accurate fit of nuclei around and above the driplines, and for applications at finite temperature.

A similar degree of accuracy is observed in the proton  root-mean-square radius (rms), displayed in  figure \ref{ArmsC} for the two extreme isotopic chains $Z=20$ and $Z=82$.
The rms of the proton distribution, $rms=\left (<r_p^2>+ S_p^2 \right)^{1/2}$, was obtained from the optimal density profile by adding the proton form factor $S_p=0.8$ fm  \cite{Marinova2013} to the  square radius defined as:
\begin{equation}\label{eq rms int}
<r_q^2>= \frac{4\pi}{N_q} \int_0^\infty r^2 \rho_q(r) r^2 dr .
\end{equation}

 A similar performance is observed along the other isotopic chains (not shown). The neutron radius, also presented in  figure \ref{ArmsC}, follows a similar trend, with larger (smaller) values than the proton ones for neutron (proton) rich systems, while close values are obtained along the stability valley.  Figure \ref{ArmsC} also shows the  rms radii using the
analytic expansion in $a_q/R_q$ that we used in equation (\ref{N fixed 2}) to obtain the radius parameters $R_q$ from the particle number conservation conditions. They are given by \cite{Mondal2016}:
\begin{equation}\label{eq rms exp}
<r_q^2> \approx {\frac{3}{5}} \left(R_q +\frac{7\pi^2}{6}\frac{a_q^2}{R_q} \right)^2.
\end{equation}
We can see that this approximate expression leads to very accurate estimates of the integrals, confirming the assumption $a_q/R_q\ll 1$ over the whole nuclear chart up to the driplines. The expansion is also very accurate when we have a background gas.

\begin{figure}
\begin{center}
\includegraphics[scale=1.]{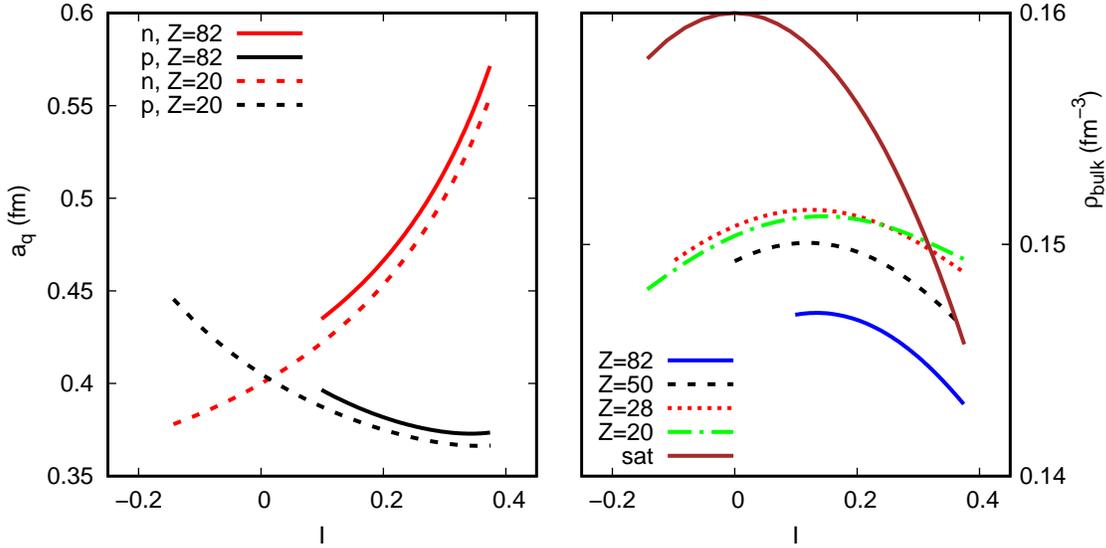}
\end{center}
\caption{Diffuseness parameters of the proton and neutron density profiles (left) and total bulk density (right)  for nuclei in the vacuum as a function of the isospin asymmetry,  for the same isotopic chains as in figures \ref{ArmsC} and \ref{aenerC}, respectively.  The saturation density of infinite nuclear matter is also shown in the same isospin range.
}
\label{anprhocC}
\end{figure}

The optimal diffusivities are displayed on the left part of figure \ref{anprhocC} for the same isotopic chains as in figure \ref{ArmsC}.
As already observed in previous works \cite{Mondal2016, Raduta2018}, the diffusivities are almost independent of the nuclear mass, but they strongly depend on the isospin, giving the most important contribution to the nuclear skin in the case of light nuclei.
 This underlines the complexity of the surface tension. Indeed, if the radius parameters are determined by the total mass numbers in a relatively trivial way (see equation (\ref{N fixed 2})), the diffuseness parameters depend in a highly non-trivial way both  on the bulk properties of matter and on the non-local  terms of the functional \cite{Aymard2016a}.

 Finally, the  total bulk density $\rho_\mathrm{bulk}=\rho_{cn}+\rho_{cp}$  for the same isotopic chains considered in figure \ref{aenerC}, is shown for completeness in the right part of  figure \ref{anprhocC}. We can see that for moderate isospin values, 
the bulk densities are systematically lower than the saturation density, as expected. The important effect of the Coulomb interaction is clearly visible for the heaviest charge $Z=82$, which suppresses the bulk density with respect to the uncharged nuclear matter expectation. Closer values 
between bulk and saturation densities are obtained towards the dripline, where
  more unbound nuclei tend to have more diluted profiles. Again, these results are in good agreement with previous ETF works \cite{Centelles1998,Warda2009,Mondal2016,Raduta2018}.  

\subsubsection{Nuclei in a gas}\label{res gas ETF}

\begin{figure}
\begin{center}
\includegraphics[scale=1.]{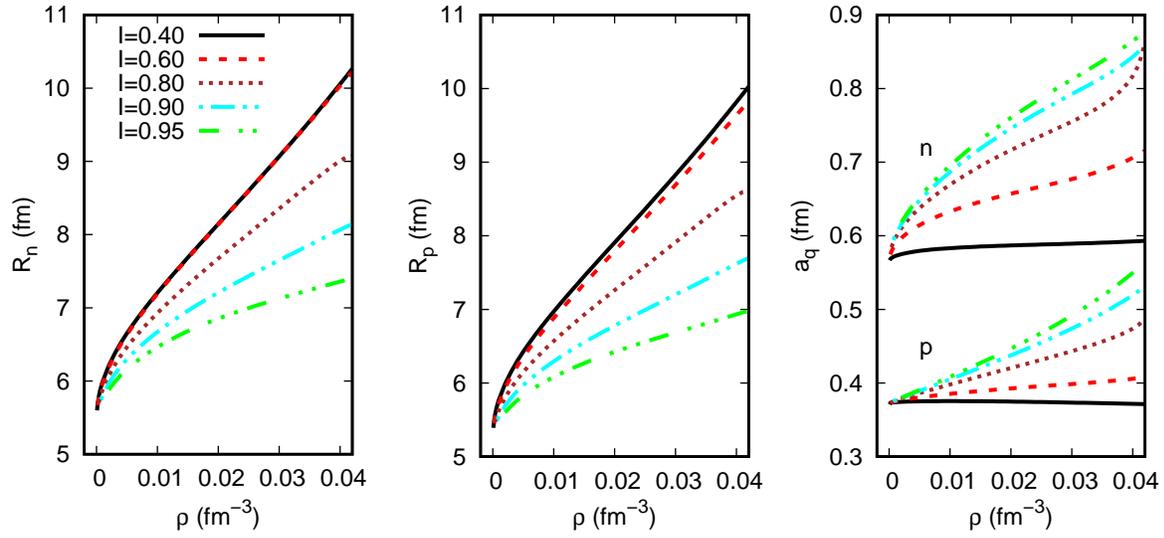}
\end{center}
\caption{Neutron radius (left panel), proton radius (central panel) and diffusivities (right panel)  from the FD profiles as a function of the total density, for different total isospins above the neutron dripline (see equation  (\ref{densities})).}
\label{aRq_aq-SLY4}
\end{figure}

\begin{figure}
\begin{center}
\includegraphics[scale=1.]{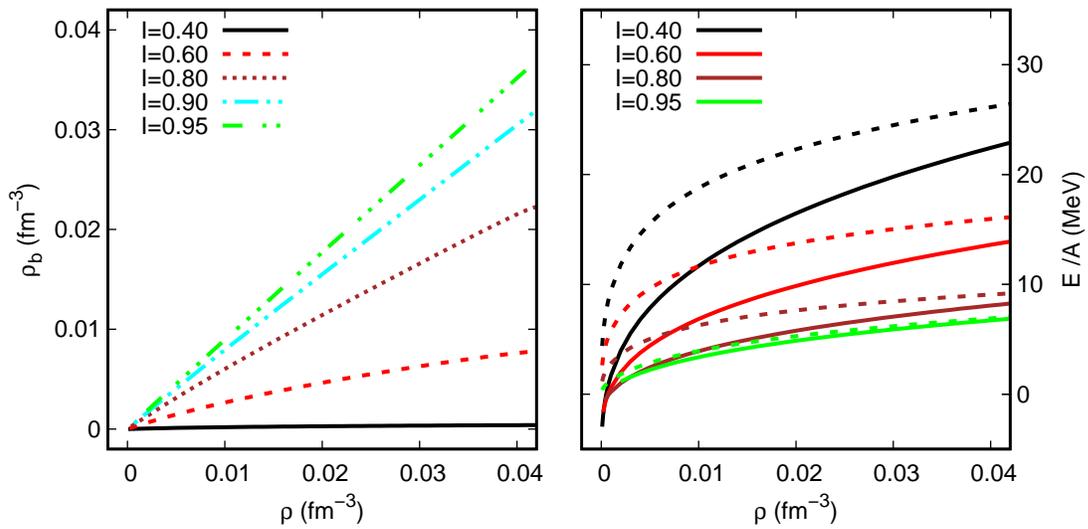}
\end{center}
\caption{ Neutron background density (left)  and total ETF energy per particle (right) as a function of the total density, for different total isospin values above the neutron dripline. The dashed lines in the right panel give the total energy of homogeneous nuclear matter with the same Sly4 functional.  }
\label{arhob_etotal-SLY4}
\end{figure}

\begin{figure}
\begin{center}
\includegraphics[scale=1.]{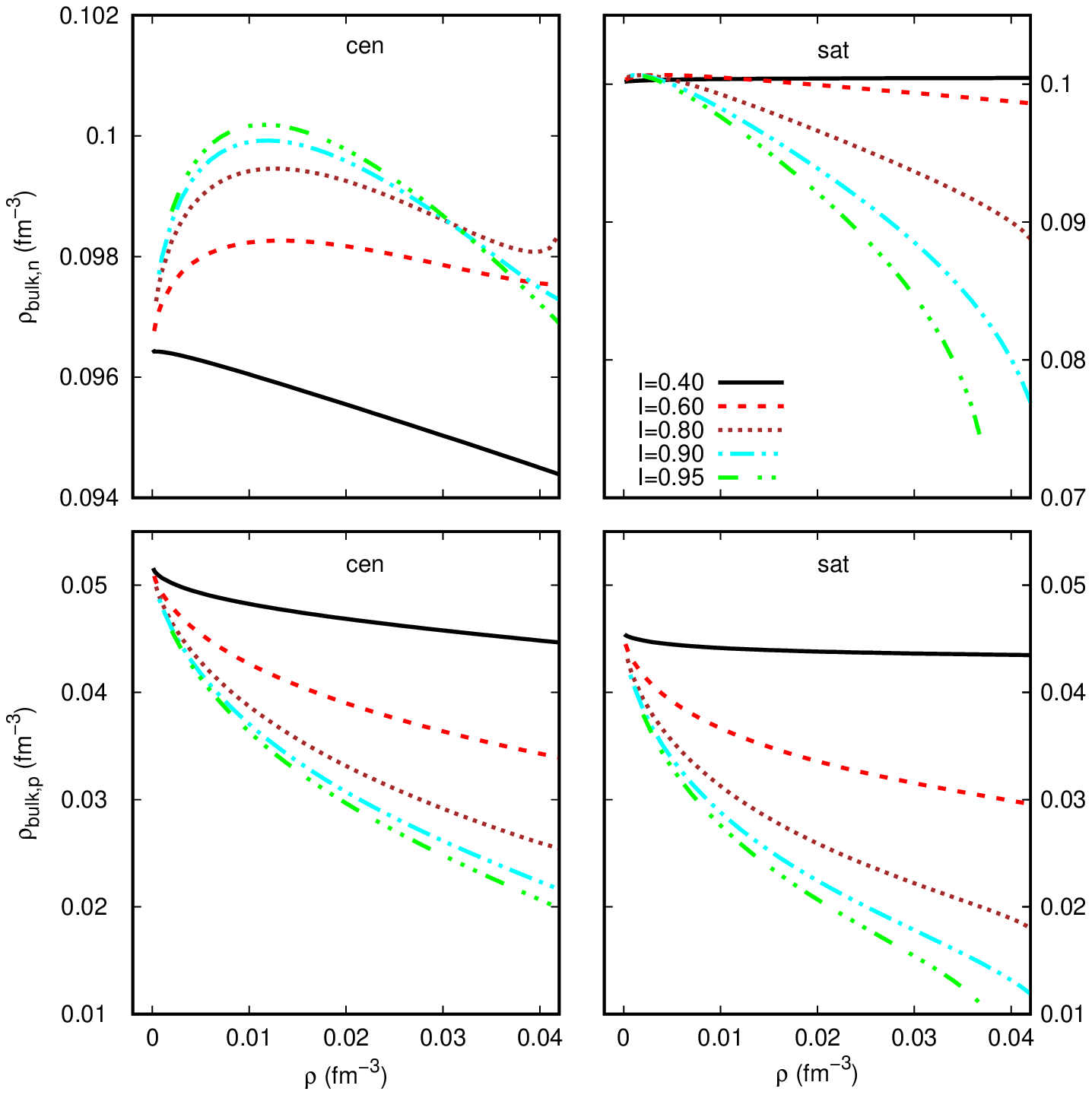}
\end{center}
\caption{ Neutron and proton central densities from equation (\ref{densities}) (left part, labelled ``cen'') and saturation densities from equation (\ref{eq:saturation}) (right part, labelled ``sat'')   as a function of the total density, for different values of the total isospin above the neutron dripline.  }
\label{arhocq-SLY4}
\end{figure}

When the isospin ratio overcomes the dripline value for a given $Z$, free neutrons naturally appear and  the parameters have to be optimized by fixing the total baryonic density.
This means that for a fixed isospin, we will not be able to independently vary the cluster mass and the density of the gas.
We will still be able to explore a large domain of masses by varying the density at fixed isospin.  

The total isospin asymmetry was varied from $I=0.00$ to $I=0.95$ in steps of 0.05. For every isospin asymmetry
 the global density  was varied from $1.\times 10^{-4}\mathrm{fm}^{-3}$ to $4.2\times 10^{-2}\mathrm{fm}^{-3}$
in steps of $1.\times 10^{-4}\mathrm{fm}^{-3}$
 until $\rho=1.\times 10^{-3}\mathrm{fm}^{-3}$, and in steps of $1.\times 10^{-3}\mathrm{fm}^{-3}$ thenceforth.
For each isospin asymmetry and global density, the system of equations was numerically solved, giving optimal values
for the parameters of the FD profiles, the neutron and proton background densities and the
WS radius within the constraints of mass conservation and charge neutrality. 
For applications in stellar matter, we will only be interested in conditions where no 
proton gas is observed, and we kept only these simulations for the following analysis. 
 For this reason, in the following we will refer to the free neutron density $\rho_{b,n}$ as ``background density'', $\rho_{b,n}\equiv
\rho_b$. 

The results for the optimal values of the variational parameters, as well as the total ETF energy, are displayed in figures \ref{aRq_aq-SLY4}, \ref{arhob_etotal-SLY4} and \ref{arhocq-SLY4}
as a function of the total density for different values of isospin in the cell. Globally, increasing density leads to larger nuclei, with more diffuse profiles  (see figure \ref{aRq_aq-SLY4}),   and to a more important contribution of the gas (see the left panel of  figure \ref{arhob_etotal-SLY4}). As we can see from the right panel of figure \ref{arhob_etotal-SLY4}, this latter feature dominates the global energetics, and the total energy per nucleon increases. This is true for all the values of $I$, but the dominance of the dripped nucleons obviously increases with the increasing isospin asymmetry. Figure \ref{arhocq-SLY4} shows that the shape of the nuclear distribution is also deeply modified with increasing density and neutron excess.  At moderate densities, the central density of the proton (neutron) distribution trivially decreases (increases) with increasing neutron excess, reflecting the change in the global baryon and proton numbers, but the trend is inverted at very high density, signalling the progressive nuclear melting in the dense medium. Note, however, the very different scales of the right panels of figure \ref{arhocq-SLY4} with respect to the left ones, showing that, despite the progressive smearing of the nuclear surface, a strong density inhomogeneity persists in the neutron distributions even at extreme isospin ratios. 

 In all figures, the thick black solid line corresponding to $I=0.4$ gives a good estimate of the behaviour just after the dripline: 
along the $I=0.4$ path, the neutron background density varies between $\rho_b=3.8\times 10^{-6}$ fm$ ^{-3}$ to $\rho_b=4.0\times 10^{-4}$ fm$ ^{-3}$ for a total mass ranging from $A=118$ to $A=637$, which corresponds to approximately 4 to 6 dripped neutrons as density increases.
 For each density, the deviations of the different quantities of the black line value  with increasing neutron excess show the modification of the nuclear shape above the dripline.
In particular we can observe from figure \ref{aRq_aq-SLY4} that while the neutron and proton radii  remain relatively constant up to a very large neutron excess, the presence of dripped nucleons strongly modifies the neutron diffusivity, already in $(\rho,I)$ configurations where the nuclei dominate over the unbound neutrons. Similar observations are in order for the
neutron and proton central densities, as shown in  figure \ref{arhocq-SLY4}. We can see that the proton 
central density rapidly decreases with both the density and isospin, but it closely follows the  value of the proton saturation density of infinite matter in this extreme isospin range, shown on the right side of  figure \ref{arhocq-SLY4}. Conversely, the very diffuse neutron density profile is associated with a relatively constant central density, which approximately corresponds to the saturation density around the drip. For extreme values of isospin, $I \ge 0.8$, the two densities are very different.

Finally, it is interesting to remark from figure \ref{arhob_etotal-SLY4} that at the most extreme value of isospin, $I=0.95$, very close to pure neutron matter, the density of the dripped nucleons is almost equivalent to the total density, and still the optimal clustered configuration is strongly energetically favoured over the homogeneous configuration, given by the dashed lines in the right part of figure \ref{arhob_etotal-SLY4}. This underlines the importance of accounting for clusters at all neutron excess.  

This ensemble of qualitative behaviours is again perfectly compatible  with previous results reported by different authors with previous ETF works \cite{Centelles1998,Warda2009,Mondal2016,Raduta2018}.  
  
\section{Parametrizing the surface energy}\label{sec surf energy}

\subsection{Formalism}\label{sec surf formalism}

For use in a CLDM \cite{Douchin2001,Carreau2019a,*Carreau2019b} or coexisting phase approximation (CPA) approach \cite{Avancini2012}, 
or for extensions to cluster distributions at finite temperature or for quasi-degenerate minima of catalyzed matter \cite{Grams2018,Barros2020}, one needs 
to evaluate the surface energy or surface tension associated with a given configuration of the WS cell
in a given thermodynamic condition specified by fixed values of the total baryonic density $\rho$, electron density $\rho_e$
(and consequently total proton fraction $Y_p=\rho_e/\rho$), and background neutron density $\rho_b$.

Following  \cite{Ravenhall1983,LS1991,Lorenz1993,newton2013,Lim2018,*Lim2019a,*Lim2019b,Carreau2019a,*Carreau2019b}, 
we choose to parametrize
the surface energy $E_\mathrm{surf}$ as:
\begin{equation}
E_\mathrm{surf}^{I} =  4\pi \sigma_\mathrm{S} R_{cl}^2 
+ 8\pi \sigma_\mathrm{C} R_{cl} ,\label{eq:esurf_newton}
\end{equation}
with a surface term and a curvature term given by:
\begin{eqnarray}
\sigma_\mathrm{S} = \sigma_0\frac{2^{p+1}+b_\mathrm{s}}{y_p^{-p} + b_\mathrm{s}+ (1-y_p)^{-p}};  \label{eq sigma surf} \\
\sigma_\mathrm{C} = \frac{\sigma_\mathrm{0c}}{\sigma_0} \alpha(\beta-y_p)\sigma_\mathrm{S}. \label{eq sigma curv}
\end{eqnarray}
Here, $R_{cl}$ is the cluster radius, $y_p$ is the cluster proton fraction, 
and $\sigma_0$, $b_\mathrm{s}$, $p$, $\sigma_\mathrm{0c}$, $\alpha$ and $\beta$ are six parameters
that need to be adjusted from the microscopic ETF calculation. The $p$ parameter is expected to
be important only for large values of isospin, more precisely when one starts 
to have a background gas \cite{Carreau2019a}, and it was kept fixed to $p=3$  in most previous works.
An extra advantage of the functional form given by equations (\ref{eq sigma surf}) and (\ref{eq sigma curv}) is that it can be straightforwardly extended to finite temperature  \cite{LS1991}. Such temperature dependence, which can be safely neglected 
for cooling applications \cite{Carreau2020}, should be taken into account for calculations at temperatures of the order of 5-10 MeV.
At these extreme temperatures, however, heavy clusters are not very important in the statistical equilibrium, which is dominated by light particles and free nucleons. We will not consider the temperature dependence in this work.
 
According to the seminal work by Ravenhall \textit{et al}. \cite{Ravenhall1983},  
$y_p$ should be the bulk cluster proton fraction, and not the proton fraction of the whole cluster.
In addition, the reference cluster density that allows us to define a cluster radius $R_{cl}$ should be the central or bulk density, 
and the cluster radius should correspond to the proton cluster radius.
In this picture, the total number of particles is not simply given by the particles of the gas plus the particles of the cluster,
 and one has an additional finite number $N_\mathrm{surf}$ of neutrons on the skin \cite{Centelles1998}.

These extra neutrons contribute to the surface energy leading to a modified expression with respect to equation (\ref{eq:esurf_newton}):
\begin{equation}
E_\mathrm{surf}^{II}= E_\mathrm{surf}^{I}
+ \mu_n N_\mathrm{surf},
\label{eq:esurf_ravenhall}
\end{equation}

\noindent where $\mu_n$ is the neutron chemical potential, which, in equilibrium, is the same
throughout the WS cell, for all cells.

The neutron chemical potential can be identified with the chemical potential of the neutron gas
\cite{lattimer1985, Grams2018}, 
\begin{equation}
\mu_n= \frac{\rmd \epsilon_\mathrm{bulk}(\rho_b)}{\rmd \rho_b}, \label{eq:mugas}
\end{equation}

\noindent where $\epsilon_\mathrm{bulk}$ is the energy density of the background neutrons, which is the nuclear bulk energy density calculated for $\rho_n=\rho_b$ and $\rho_p=0$.

In order to extract the surface energy parameters from the ETF calculation, we introduce the standard  decomposition of the Wigner Seitz energy as:

\begin{equation}
E_\mathrm{ETF}= E_{\mathrm{bulk,nuc}} +E_{\mathrm{bulk,Coul}} +E_\mathrm{surf}, \label{eq dec}
\end{equation} 

\noindent where the total density $\rho=A/V_\mathrm{WS}$ and proton fraction $Y_p=Z/A$ are imposed, 
and the Wigner-Seitz volume $V_\mathrm{WS}$ together with the density profiles are variationally determined
assuming  a given nuclear functional (see section \ref{formalism}).
The surface energy will be defined as the subtraction of the bulk (nuclear and Coulomb) terms from the energy as obtained by the microscopic ETF calculation, $E_\mathrm{surf}\equiv E_\mathrm{ETF}-E_{\mathrm{bulk,nuc}} -E_{\mathrm{bulk,Coul}} $.
 We have therefore to specify the bulk terms. 

Defining the bulk kinetic energy density as:
\begin{equation}
\mathcal {H}_{\mathrm{bulk,kin}}(\rho_n,\rho_p)=\sum_{q=n,p} \frac{\hbar^2}{2 m_q^*} \tau_{0q} ,
\end{equation}
one has for nuclei in the vacuum:

\beqaa 
E_{\mathrm{bulk,nuc}}= \left. \left(  \mathcal {H}_{\mathrm{bulk,kin}}   + v  \right) \right|_{\rho_{cl}}
V_{cl};  \\
E_{\mathrm{bulk,Coul}}= \frac{e^2}{4\pi}\frac{3Z^2}{5R_{cl}}
-\frac{3e^2}{16 \pi} \left(\frac{3}{\pi} \right)^{1/3}\rho_{p,cl}^{4/3}V_{cl},
\label{eq bulk vacuum}
\eeqaa 

\noindent where $V_{cl}=(4/3)\pi R_{cl}^3$ is the volume of the homogeneous sphere defined by a 
constant  density $\rho_{cl}$, to be defined below together with 
the cluster radius $R_{cl}$.

For nuclei in a gas, one has:
\beqa
E_{\mathrm{bulk,nuc}} = \left. \left( \mathcal {H}_{\mathrm{bulk,kin}}  + v  \right) \right|_{\rho_{cl}}
V_{cl} + \left. \left(\mathcal {H}_{\mathrm{bulk,kin}}  + v  \right) \right|_{\rho_{b}}
(V_\mathrm{WS}-V_{cl}).
\label{eq bulk gas}
\eeqa
The last term accounts for the energy of the gas.

We remark that to represent a bulk term, the Coulomb energy is calculated 
 for the simple density profile of constant density in the cluster and in the background gas. This means that what we call ``surface energy'' 
will contain both nuclear and Coulomb contributions due to the presence of an interface between the nucleus and the background gas.

For $\rho_{bp}=0$, using
$Z=(4/3)\pi R_{cl,p}^3 \rho_{cl,p}=(4/3)\pi R_\mathrm{WS}^3 \rho_e$, one finds:
\beqaa 
E_{\mathrm{bulk,Coul}}=& \frac{3}{5} \frac{e^2}{4\pi} \frac{Z^2}{R_{cl,p}}
\left( 1-\frac{3}{2} \frac{R_{cl,p}}{R_\mathrm{WS}}
+\frac{1}{2} \frac{R_{cl,p}^3}{R_\mathrm{WS}^3} \right) \nonumber \\
& -\frac{3e^2}{16 \pi} \left(\frac{3}{\pi} \right)^{1/3} Z
(\rho_{cl,p}^{1/3} + \rho_e^{1/3}). \label{ex coulomb uniform 2}
\eeqaa 
Details are given in \ref{appC}.

To have a complete closed set of equations, we need to specify the effective cluster densities $\rho_{cl},\rho_{cl,p}$ and cluster radii $R_{cl},R_{cl,p}$. 
Concerning the radii, the simplest prescription consists in ignoring the presence of neutrons in the skin, $N_\mathrm{surf}=0$. 
Within this scheme, the surface energy is given by equation (\ref{eq:esurf_newton}).
Then a single radius is needed, $R_{cl,n}=R_{cl,p}=R_{cl}$, which can be determined from the charge conservation condition in the cell, as a function of the proton cluster density  $\rho_{cl,p}$ (recall that there are no background protons, $\rho_{bp}=0$):

\beqaa 
Z\equiv Z_{cl}= \frac{4\pi}{3} \rho_{cl,p}R_{cl,p}^3
.  \label{eq Rp def}
\eeqaa 
 
The neutron number conservation then allows determining the neutron density of the cluster $\rho_{cl,n}$, as well as the cluster neutron number $N_{cl}$ and the cluster proton fraction $y_p=Z_{cl}/(Z_{cl}+N_{cl})$ via:  

\begin{eqnarray}
N_{cl}= \frac{4\pi}{3} \rho_{cl,n}R_{cl,p}^3 , \label{eq Ncl def} \\
N= \frac{4\pi}{3} \left[\rho_{cl,n}R_{cl,p}^3 +\rho_{b}(R_\mathrm{WS}^3 - R_{cl,p}^3)\right] ,  \label{eq rhon def}
\end{eqnarray}

\noindent where $Z, N$, $\rho_{b}$ and $R_\mathrm{WS}$ come from the ETF calculation.

The last quantities to be specified to close the system of equations are the cluster densities $\rho_{cl}$, $\rho_{cl,p}$.
Different options exist in the literature  for the cluster density $\rho_{cl}$, and they do not necessarily lead to the same definitions and behaviours for the surface energy and its isospin dependence \cite{Aymard2014}.
Since this density is the one that defines the bulk quantities, it should correspond to a local quantity and not to an   average over the spatial extension of the cell. The most natural choice is then to employ  the central density of the
density profile (referred to as ``central density'') \cite{Ravenhall1983}. Another option, introduced in  \cite{papakonstantinou2013}, consists in using the saturation density corresponding to the cluster isospin asymmetry,
referred to as ``saturation density''. In fact, this density is the solution of the ETF variational equations in the bulk limit, that is, in slab geometry for $z\to -\infty$ \cite{treiner1986}. 
 
Performing a Taylor expansion at second order in the asymmetry $I$, 
the saturation density is given by \cite{papakonstantinou2013}:
\begin{equation}
\rho_{cl}(I) = \rho_\mathrm{sat}(0) \left( 1 - \frac{3 L_\mathrm{sym} I^2}{K_\mathrm{sat}+K_\mathrm{sym} I^2} \right).
\label{eq_asym_rho0}
\end{equation}
In this expression, 
$K_\mathrm{sat}=9\rho_\mathrm{sat}^2\partial^2(\mathcal{H}/\rho)/\partial\rho^2|_{\rho_\mathrm{sat}}$ is the nuclear (symmetric) matter incompressibility,
and
$L_\mathrm{sym}=3\rho_\mathrm{sat}\partial (\mathcal{H}_\mathrm{sym}/\rho)/\partial \rho |_{\rho_\mathrm{sat}}$  and 
$K_\mathrm{sym}=9\rho_\mathrm{sat}^2\partial^2 (\mathcal{H}_\mathrm{sym}/\rho)/\partial \rho^2 |_{\rho_\mathrm{sat}}$ 
are the slope and curvature of the symmetry energy at (symmetric) saturation, where 
we have introduced the usual definition of the symmetry energy density :
\begin{equation}
\mathcal{H}_\mathrm{sym}(\rho)= \frac{\rho^{2}}{2} \left.\frac{\partial^2 \mathcal{H}_\mathrm{ETF}}{\partial (\rho \delta)^{2}}\right|_{\delta=0}.
\end{equation}
For better precision at high asymmetry, one can also use an improved approximation keeping third and fourth order terms in the expansion. The saturation density is then given by the solution of the following equation:
\beqaa \label{eq:saturation}
x^3 \left(\frac{Z_\mathrm{sat}}{6}+I^2\frac{Z_\mathrm{sym}}{6}\right) & +
x^2 \left(\frac{Q_\mathrm{sat}}{2}+I^2\frac{Q_\mathrm{sym}}{2}\right) \nonumber \\
&+ x (K_\mathrm{sat}+I^2 K_\mathrm{sym}) +I^2 L_\mathrm{sym} =0 ,
\eeqaa 
where $x=(\rho-\rho_\mathrm{sat})/3\rho_\mathrm{sat}$,
and the higher order EOS empirical  parameters $Q_{\mathrm{sat(sym)}}=27\rho_\mathrm{sat}^3\partial^3 (\mathcal{H}_{\mathrm{(sym)}}/\rho)/\partial \rho^3 |_{\rho_\mathrm{sat}}$ and $Z_{\mathrm{sat(sym)}}=81\rho_\mathrm{sat}^4\partial^4 (\mathcal{H}_{\mathrm{(sym)}}/\rho)/\partial \rho^4 |_{\rho_\mathrm{sat}}$
have to be specified from the chosen energy functional.

Whatever the order of the expansion, 
 the saturation density is evaluated at the isospin asymmetry of the cluster given by  $I_{cl}=(N_{cl}-Z_{cl})/(N_{cl}+Z_{cl})$,
and it is related to the proton and neutron cluster densities entering  equations (\ref{eq Rp def})-(\ref{eq rhon def}) by
$\rho_{cl}=\rho_{cl,p}+\rho_{cl,n}$.

 If we associate the cluster density $\rho_{cl}$ required to calculate the cluster radius $R_{cl}$ and the corresponding cluster volume $V_{cl}$ to the saturation density equation (\ref{eq_asym_rho0}), this will provide a fully analytical surface energy functional, 
which can also be used when no variational calculation is done, and the optimal density profile is not known. 
For this reason, a parametrization employing the saturation density would be of more practical use in EOS for astrophysical applications \cite{Raduta2018}, provided the associated surface tension reproduces with sufficient accuracy the microscopic theory. We will verify this point in the next section.

An alternative prescription consists in employing the central density as a definition of the effective cluster density.
This quantity is directly extracted 
from the microscopic calculation as $\rho_{cl,q}=\rho_q(r=0)$, where $\rho_q(r)$ is the ETF density profile from equation (\ref{densities}).  
 If this prescription is employed, the inclusion of neutrons in the skin cannot be avoided.
Indeed, independent definitions of $\rho_{cl,n}$ and $\rho_{cl,p}$ are not compatible with the simultaneous validity 
of equations (\ref{eq Rp def}) and (\ref{eq rhon def}), or, in other words, they are not compatible with the definition of a single radius for the protons and neutrons. Baryon number conservation now implies:

\beqa
R_{cl,n}^3= \frac{\rho-\rho_e-\rho_{b}}{\rho_{cl,n}-\rho_{b}} R_\mathrm{WS}^3 , \label{eq rcln}
\eeqa

\noindent leading to a finite number of neutrons on the interface:

\beqaa 
N_\mathrm{surf}&= N -\frac{4\pi}{3} [\rho_{cl,n}R_{cl,p}^3 +\rho_{b}(R_\mathrm{WS}^3 - R_{cl,p}^3)] \nonumber \\
&= \frac{4\pi}{3} [(\rho_{cl,n} -\rho_{b})(R_{cl,n}^3 - R_{cl,p}^3)].
\label{n numb skin}
\eeqaa  

In this scheme, the surface energy is given by equation (\ref{eq:esurf_ravenhall}). The cluster radii are given by equations (\ref{eq rcln}) and (\ref{eq Rp def}) as a function of the two input densities $\rho_{cl,p}$ and $\rho_{cl,n}$. To have a consistent evaluation of the  cluster proton fraction $y_p=Z_{cl}/A_{cl}=\rho_{cl,p}/\rho_{cl}$, and compatibility with the calculation of the bulk Coulomb energy, we necessarily need to define the cluster volume from the proton radius, $V_{cl}=(4/3)\pi R_{cl,p}^3$. We remark that this also amounts to defining the cluster neutron number from the same radius,  as in the case of the saturation density choice:  $N_{cl}=(4/3)\pi \rho_{cl,n}R_{cl,p}^3$, with excess neutrons being considered as skin neutrons.

Of course, it is also possible to consider the surface neutrons and the two different cluster radius equations (\ref{eq rcln}) and
(\ref{n numb skin}) in the case where the cluster density is defined as the saturation density from equation (\ref{eq_asym_rho0}).
In this case, to close the system of equations, we additionally need to write $\rho_{cl,p}=\rho_{cl}y_p$.

In the following, we will take the prescription in which the cluster density is given by the saturation density, and $N_\mathrm{surf}=0$, as our first choice. With this prescription, we stick to the simplest approach where the cluster radius and particle numbers are given by 
equations (\ref{eq Rp def}) and (\ref{eq rhon def}).  The surface energy, given by  equation (\ref{eq:esurf_newton}), will be fitted from the ETF calculation using equation (\ref{eq dec}). 

The results will be compared to the more sophisticated prescription where the cluster densities are directly extracted from the ETF calculation as central densities,  neutrons on the interface are accounted for with equation (\ref{n numb skin}), and the surface energy is given by equation (\ref{eq:esurf_ravenhall}). Though the total energy does not vary, as it is in both cases taken from the ETF calculation, the relative weight of what is defined as ``bulk'' and ``surface'' will obviously be different with the two prescriptions. 

We will see that  an excellent fit can be obtained using a single radius and the saturation density, thus leading to a fully analytical prescription for the surface energy as a function of the particle numbers, $E_\mathrm{surf}=E_\mathrm{surf}(N_{cl},Z_{cl})$. 
 
\begin{figure}
\begin{center}
\includegraphics[scale=1.]{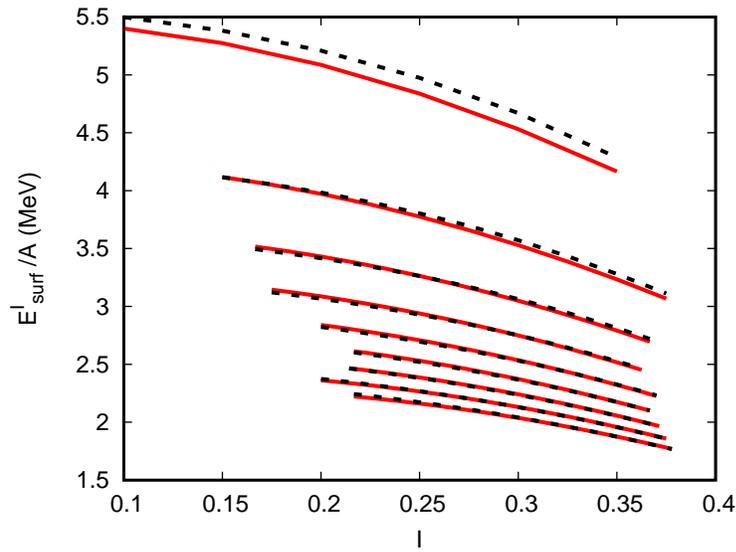}
\end{center}
\caption{Best fit of the surface energy 
as a function of the isospin $I$, for different  mass numbers starting at $A=40$ (highest line) and increased by steps of $\Delta A=40$ up to $A=360$ (lowest line). 
Solid lines: ETF results; dashed lines: fit with equation (\ref{eq:esurf_newton}).
The saturation density is used to estimate the cluster density (see text for more details).
}
\label{aesurfsatCV-SLY4}
\end{figure}

\begin{figure}                                               
\begin{center}                                               
\includegraphics[scale=1.]{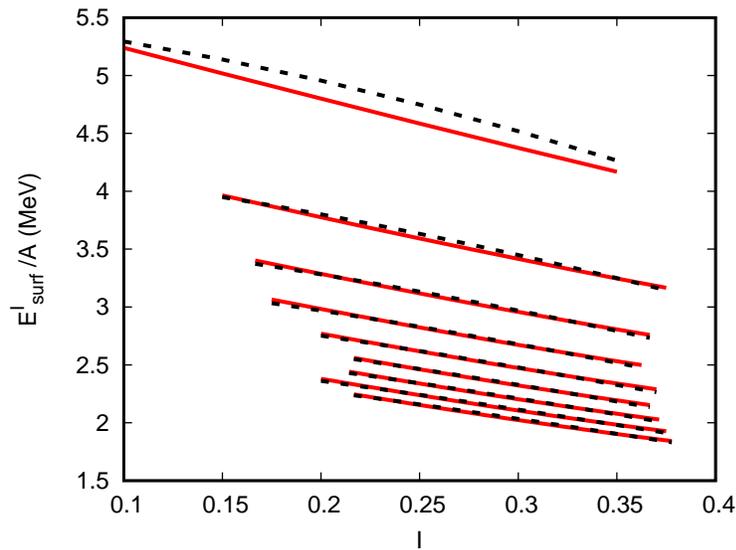}                 
\end{center}
\caption{Same as figure  \ref{aesurfsatCV-SLY4}, but neutrons in the skin are allowed and 
the central density  is used to estimate the cluster density (see text for more details).
 }
\label{aesurfcenPCV-SLY4}
\end{figure}

 Different applications can be foreseen for this parametrized surface energy.
A first possible application concerns equations of state based on the  liquid drop model and the evaluation of cluster distributions at finite temperature \cite{Carreau2020}. In this case, for a given thermodynamic condition, all cluster particle numbers  $Z_{cl}$, $N_{cl}$ are considered. The total cluster density $\rho_{cl}$  can then be taken from equation (\ref{eq_asym_rho0}), and  equations (\ref{eq Rp def}) and (\ref{eq Ncl def})  can be used  to obtain the partial densities $\rho_{cl,q}$ and radius $R_{cl}$, to be used in equation (\ref{eq:esurf_newton}).
 
Another potential application concerns equations of state in the single nucleus or CPA approach \cite{Avancini2012}.
In this case, for a given thermodynamic condition $(\rho,Y_p)$,  the equilibrium equations provide the cluster densities 
$\rho_{cl,p}$ and $\rho_{cl,n}$ and  proton fraction $y_p=\rho_{cl,p}/(\rho_{cl,n}+\rho_{cl,p})$. 
A single radius is assumed for the cluster in this approximation, which should be variationally determined from the 
competition between the Coulomb and surface energies. 
The variational equation reads:
\begin{equation}
0=\frac{\rmd ~~}{\rmd R_{cl}} \frac{E_{\mathrm{bulk,Coul}}+E_\mathrm{surf}^I}{R_{cl}^3},
\end{equation}
which reduces to the well-known Baym virial theorem \cite{BBP} if only the first  leading terms in $R_{cl}$ are retained in both Coulomb and surface terms:  
\begin{equation}
E_\mathrm{surf}=2E_{\mathrm{bulk,Coul}} \ .
\end{equation}

\begin{figure}
\begin{center}
\includegraphics[scale=1.]{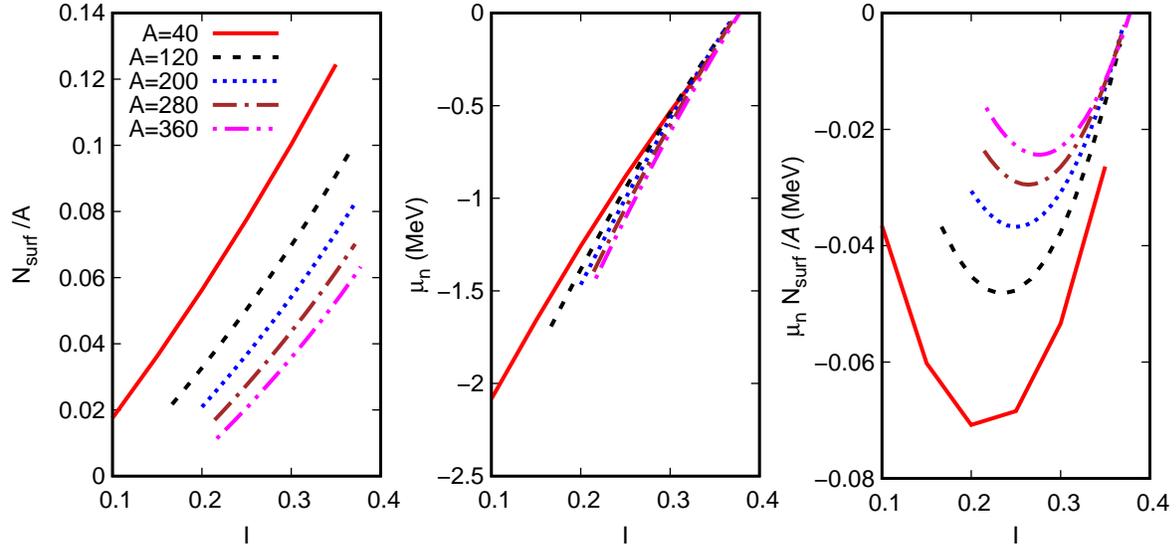}
\end{center}
\caption{ As a function of isospin for different mass numbers from $A=40$ (highest line in the  left and central panels, lowest line in the right panel) and increased by steps of $\Delta A=40$, we show for the central density condition: 
  the fraction of neutrons in the skin $N_\mathrm{surf}/A$ (left), the neutron chemical potential 
 $\mu_n$ (central), their product $\mu_n N_\mathrm{surf}/A$ (right).
}
\label{auNsurfcenPCV-SLY4}
\end{figure}

\subsection{Results}\label{sec surf results}

\subsubsection{Results: Nuclei in the vacuum}\label{res vacuum}

ETF calculations were performed for nuclei in the range of $A=40$ to $A=360$, in steps of $\Delta A=10$. 
Since we are only interested in the neutron rich side, the isospin $I=(N-Z)/A$ was varied from the value corresponding to the
most stable isotope for each $A$ \cite{Audi2017}, up to the value corresponding to the neutron dripline, evaluated from the condition of a vanishing neutron chemical potential, $\mu_n=0$.  For $A>295$, in the absence of experimental information, we kept as minimal $I$ value the one corresponding to $A=295$.
 
In principle, five parameters have to be fitted, namely  $\sigma_0$, $b_\mathrm{s}$, $p$, $\beta$, and the product
$\alpha\sigma_\mathrm{0c}$, for which we take $\alpha=5.5$ fm as in \cite{newton2013}.
However, it was observed in \cite{Carreau2019a,*Carreau2019b} that the adimensional $p$ parameter is crucial at extreme isospin, such as the one encountered at the crust-core transition of neutron stars, while it is a redundant parameter below the dripline. For this reason, we fixed this parameter to an arbitrary value from $p=0.5$ to $p=5$, and fitted the other parameters using the subroutine MRQMIN from  \cite{numerical_recipes1997}.

 The quality of the fit is estimated from the $\chi^2$ defined as:
\begin{equation}
\chi^2 =\frac{1}{N-N_{\mathrm{par}}-1} \sum_{i=1}^N \frac{((E/A)_{\mathrm{fit},i}-(E/A)_{\mathrm{ETF},i})^2 }{ (\sigma/A_i)^2},
\end{equation}
where $N$ is the total number of nuclei included in 
the fit, $N_{\mathrm{par}}=4$ is the number of fit parameters, and we take $\sigma=2$ MeV as the average  precision of the theoretical formula to represent the microscopic ETF ``data'',
determined such as to have $\chi_2^{min}\approx 1$ considering
the best estimate of the parameters for the optimal fit of the larger data set including the neutron gas (see next section).

The ETF results  fitted using equation (\ref{eq:esurf_newton}), and estimating the cluster density via the saturation density equation (\ref{eq_asym_rho0}), are shown in figure  \ref{aesurfsatCV-SLY4}, while the results of the fit allowing neutrons in the skin through equation (\ref{eq:esurf_ravenhall}) are displayed in figure \ref{aesurfcenPCV-SLY4}.

 In both cases, the fit procedure indicated that $\beta \rightarrow \infty$, but $\beta \cdot \sigma_\mathrm{0c}$ is constant. In other words, the variational results imposed that the dependence of the curvature term must be the same as the surface term in equation (\ref{eq:esurf_newton}). We therefore imposed for the curvature term in equation (\ref{eq:esurf_newton}):
\begin{equation}
\sigma_\mathrm{C}=\frac{\sigma_\mathrm{0c}}{\sigma_0}\alpha\sigma_\mathrm{S}.\label{eq:curv}
\end{equation}
The same prescription was adopted in the presence of a neutron gas. $N_{\mathrm{par}}=3$ now.

In all cases, the fit procedure converged rapidly after only a few iterations, and 
we can see from figures \ref{aesurfsatCV-SLY4} and \ref{aesurfcenPCV-SLY4} that both prescriptions lead to a satisfactory 
reproduction of the microscopic results. The results displayed in the figures correspond to the value of the $p$ parameter leading to the best fit in each case, but fits of comparable $\chi^2$ are obtained for a large interval of $p$, as it is shown in tables  \ref{tpBC-sat} and \ref{tpBC-cen}.

\begin{table}
\lineup
\caption{\label{tpBC-sat} Parameters fitted for nuclei in the vacuum and quality of the corresponding fit for different choices of the $p$ parameter. Saturation density was employed (see text).
 604 nuclei were considered. The last column gives the number of iterations needed to achieve convergence.}
\begin{indented}
\item[]\begin{tabular}{@{}cccccc}
  \br
  $b_\mathrm{s}$ & $\sigma_0$(MeV$\cdot$fm$^ {-2}$) & $\sigma_\mathrm{0c}$(MeV$\cdot$fm$ ^{-2}$) & $p$ & $\chi^2$  & iter \\
  \mr
\0\0  1.044~ &  0.99249~ &  0.070321~  & 2.0~ &  2.2565~ &    5\\
\0\0  2.101~ &  0.99064~ &  0.070497~  & 2.1~ &  2.1371~ &    4\\
\0\0  3.371~ &  0.98876~ &  0.070675~  & 2.2~ &  2.0219~ &    4\\
\0\0  4.891~ &  0.98687~ &  0.070855~  & 2.3~ &  1.9111~ &    4\\
\0\0  6.698~ &  0.98496~ &  0.071038~  & 2.4~ &  1.8051~ &    4\\
\0\0  8.835~ &  0.98304~ &  0.071223~  & 2.5~ &  1.7043~ &    4\\
\0 11.355~ &  0.98110~ &  0.071409~  & 2.6~   &  1.6090~ &    4\\
\0 14.313~ &  0.97915~ &  0.071598~  & 2.7~   &  1.5193~ &    4\\
\0 17.775~ &  0.97719~ &  0.071788~  & 2.8~   &  1.4358~ &    4\\
\0 21.815~ &  0.97522~ &  0.071981~  & 2.9~   &  1.3585~ &    4\\
\0 26.518~ &  0.97324~ &  0.072174~  & 3.0~   &  1.2879~ &    4\\
\0 31.979~ &  0.97125~ &  0.072370~  & 3.1~   &  1.2242~ &    4\\
\0 38.308~ &  0.96925~ &  0.072567~  & 3.2~   &  1.1677~ &    4\\
\0 45.629~ &  0.96725~ &  0.072765~  & 3.3~   &  1.1186~ &    4\\
\0 54.081~ &  0.96525~ &  0.072964~  & 3.4~   &  1.0772~ &    4\\
\0 63.826~ &  0.96324~ &  0.073165~  & 3.5~   &  1.0437~ &    4\\
\0 75.043~ &  0.96123~ &  0.073367~  & 3.6~   &  1.0184~ &    4\\
\0 87.939~ &  0.95921~ &  0.073570~  & 3.7~   &  1.0014~ &    4\\
102.745~ &  0.95720~ &  0.073773~  & 3.8~     &  0.9930~ &    4\\
119.726~ &  0.95518~ &  0.073978~  & 3.9~     &  0.9933~ &    4\\
139.180~ &  0.95317~ &  0.074183~  & 4.0~     &  1.0026~ &    4\\
161.445~ &  0.95116~ &  0.074390~  & 4.1~     &  1.0210~ &    4\\
186.905~ &  0.94915~ &  0.074596~  & 4.2~     &  1.0487~ &    4\\
215.993~ &  0.94715~ &  0.074804~  & 4.3~     &  1.0858~ &    4\\
249.198~ &  0.94515~ &  0.075012~  & 4.4~     &  1.1326~ &    4\\
287.075~ &  0.94315~ &  0.075220~  & 4.5~     &  1.1891~ &    4\\
330.252~ &  0.94116~ &  0.075429~  & 4.6~     &  1.2554~ &    4\\
379.437~ &  0.93918~ &  0.075638~  & 4.7~     &  1.3317~ &    4\\
435.432~ &  0.93720~ &  0.075847~  & 4.8~     &  1.4181~ &    4\\
499.141~ &  0.93524~ &  0.076056~  & 4.9~     &  1.5147~ &    4\\
571.588~ &  0.93328~ &  0.076266~  & 5.0~     &  1.6215~ &    4\\
\br
\end{tabular}
\end{indented}
\end{table}

\begin{table}
\lineup
\caption{\label{tpBC-cen} Parameters fitted for nuclei in the vacuum and quality of the corresponding fit for different choices of the $p$ parameter. Central density was employed (see text). 604 nuclei were considered.
The last column gives the number of iterations needed to achieve convergence.}
\begin{indented}
\item[]\begin{tabular}{@{}cccccc}
  \br
  $b_\mathrm{s}$ & $\sigma_0$(MeV$\cdot$ fm$ ^{-2}$) & $\sigma_\mathrm{0c}$(MeV$\cdot$fm$ ^{-2}$) & $p$ & $\chi^2$  & iter \\
  \mr
\0 -2.431~ &  0.93252~ &  0.058036~  & 0.5~   &  1.3492~ &    6 \\
\0 -2.482~ &  0.93177~ &  0.058111~  & 0.6~   &  1.3789~ &    4 \\
\0 -2.513~ &  0.93101~ &  0.058188~  & 0.7~   &  1.4102~ &    3 \\
\0 -2.520~ &  0.93022~ &  0.058267~  & 0.8~   &  1.4433~ &    3 \\
\0 -2.497~ &  0.92942~ &  0.058348~  & 0.9~   &  1.4782~ &    3 \\
\0 -2.438~ &  0.92859~ &  0.058432~  & 1.0~   &  1.5149~ &    4 \\
\0 -2.336~ &  0.92775~ &  0.058517~  & 1.1~   &  1.5535~ &    4 \\
\0 -2.182~ &  0.92689~ &  0.058605~  & 1.2~   &  1.5941~ &    4 \\
\0 -1.967~ &  0.92601~ &  0.058694~  & 1.3~   &  1.6366~ &    4 \\
\0 -1.681~ &  0.92512~ &  0.058786~  & 1.4~   &  1.6813~ &    4 \\
\0 -1.310~ &  0.92420~ &  0.058879~  & 1.5~   &  1.7280~ &    4 \\
\0 -0.840~ &  0.92327~ &  0.058974~  & 1.6~   &  1.7770~ &    4 \\
\0 -0.256~ &  0.92233~ &  0.059071~  & 1.7~   &  1.8281~ &    4 \\
\0\m  0.462~ &  0.92137~ &  0.059170~  & 1.8~ &  1.8816~ &    4 \\
\0\m  1.336~ &  0.92040~ &  0.059271~  & 1.9~ &  1.9374~ &    4 \\
\0\m  2.388~ &  0.91941~ &  0.059374~  & 2.0~ &  1.9956~ &    4 \\
\0\m  3.647~ &  0.91841~ &  0.059478~  & 2.1~ &  2.0562~ &    4 \\
\0\m  5.144~ &  0.91740~ &  0.059584~  & 2.2~ &  2.1194~ &    4 \\
\0\m  6.915~ &  0.91637~ &  0.059691~  & 2.3~ &  2.1851~ &    4 \\
\0\m  9.001~ &  0.91534~ &  0.059800~  & 2.4~ &  2.2534~ &    4 \\
\m 11.447~ &  0.91429~ &  0.059910~  & 2.5~   &  2.3243~ &    4 \\
\m 14.305~ &  0.91323~ &  0.060022~  & 2.6~   &  2.3980~ &    4 \\
\m 17.635~ &  0.91216~ &  0.060136~  & 2.7~   &  2.4744~ &    4 \\
\m 21.504~ &  0.91108~ &  0.060250~  & 2.8~   &  2.5535~ &    4 \\
\m 25.987~ &  0.90999~ &  0.060366~  & 2.9~   &  2.6355~ &    4 \\
\m 31.170~ &  0.90890~ &  0.060484~  & 3.0~   &  2.7203~ &    4 \\
\m 37.150~ &  0.90779~ &  0.060602~  & 3.1~   &  2.8080~ &    4 \\
\m 44.037~ &  0.90668~ &  0.060722~  & 3.2~   &  2.8986~ &    4 \\
\m 51.953~ &  0.90556~ &  0.060843~  & 3.3~   &  2.9922~ &    4 \\
\m 61.040~ &  0.90444~ &  0.060965~  & 3.4~   &  3.0888~ &    4 \\
\m 71.454~ &  0.90331~ &  0.061088~  & 3.5~   &  3.1884~ &    4 \\
\br
\end{tabular}
\end{indented}
\end{table}

These tables show that there is an anticorrelation between the $\sigma_0$ parameter, corresponding to the surface tension of symmetric nuclei, and the $b_\mathrm{s}$ one, governing the isospin dependence for moderate values of isospins \cite{newton2013}. This anticorrelation was already observed in  \cite{Carreau2019a,*Carreau2019b} on the fit of experimental data. We additionally observe a correlation between the surface tension parameter $\sigma_0$ and the curvature parameter $\sigma_\mathrm{0c}$.  

In the case of the fit using the central density, as discussed before, the neutron radius does not coincide with the proton radius and neutrons can appear on the surface of the nucleus, modifying the global energetics according to equation (\ref{eq:esurf_ravenhall}).
In spite of that, we can see that the numerical values of the surface energy in figure \ref{aesurfcenPCV-SLY4} are only slightly reduced 
with respect to the ones of figure \ref{aesurfsatCV-SLY4} obtained with a single radius. Even the isospin dependence is almost unaffected by the account of neutrons in the skin: only for the lightest nuclei the decrease of the surface energy with increasing isospin is slightly steeper, but this effect is not properly accounted for by the fit.

To better understand this behaviour, 
the effect of the skin is further explored  in figure  \ref{auNsurfcenPCV-SLY4}, which displays the behaviour as a function of the isospin 
of the neutron chemical potential and surface neutrons, as extracted from the fit of the ETF results using the decomposition between bulk and surface energies of equation (\ref{eq:esurf_ravenhall}).
The chemical potential was directly extracted from the optimal ETF profile as: 
\beqaa
\mu_{n}=\left. \frac{\partial E_\mathrm{ETF}}{\partial N_n} \right|_{V_\mathrm{WS},N_p} =&
\frac{\partial E_\mathrm{ETF}}{\partial a_n} \frac{\partial a_n}{\partial N_n} +
\frac{\partial E_\mathrm{ETF}}{\partial \rho_{cn}} \frac{\partial \rho_{cn}}{\partial N_n}
\nonumber \\
&+ \frac{\partial E_\mathrm{ETF}}{\partial \rho_{bn}} \frac{\partial \rho_{bn}}{\partial N_n}
+ \frac{\partial E_\mathrm{ETF}}{\partial R_n} \frac{\partial R_n}{\partial N_n} .
\eeqaa
We can see that $\mu_n$  increases monotonically to reach zero at the dripline, and it shows small finite size effects, namely a moderate decrease with $A$ at fixed $I$, as it can be expected for a bulk quantity. The number of surface neutrons also monotonically increases with the isospin, as expected. This surface quantity approximately scales with the area of the interface $\propto A_{cl}^{2/3}$ and, therefore, it gives a greater contribution to the total energy in the case of lighter nuclei.  

\begin{figure}
\begin{center}
\includegraphics[scale=1.]{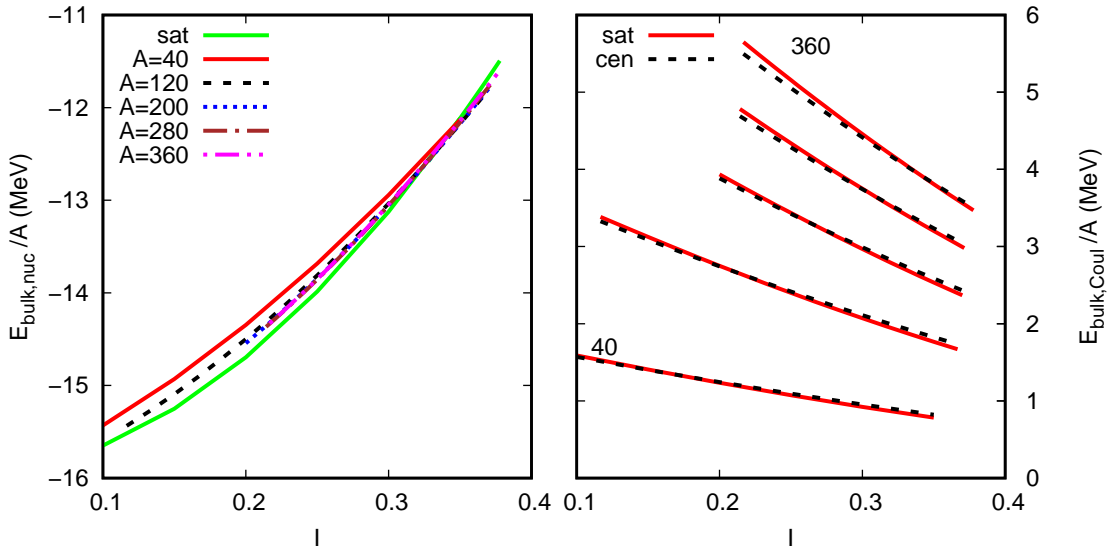}
\end{center}
\caption{Bulk nuclear (left panel) and Coulomb (right panel) energies per nucleon for nuclei in the vacuum as a function of the total isospin $I$ for different mass numbers $A$. Both prescriptions for the bulk density (saturation and central density) are presented.    
}
\label{aebulk_coulCV-SLY4}
\end{figure}

Because of  the negative sign of the chemical potential, the product of the two quantities shows a characteristic minimum at a value of $I$ which increases with $A$, and which results from the competition between the attraction of the nuclear mean field and the increasing neutron excess. Because of this compensation, the energetic contribution of the skin neutrons is negligible for all isospin asymmetries, and  the two prescriptions  lead to very similar surface energies and equivalently good representations of the global ETF energetics. The  difference in the surface energy observed  between figures \ref{aesurfsatCV-SLY4} and \ref{aesurfcenPCV-SLY4} can therefore not be ascribed to the inclusion (or not) of skin neutrons. Instead, this difference, which is only sizeable for moderate $I$ and small $A$, can essentially be explained by the different prescriptions for the bulk density adopted in the two figures, which modifies the bulk energy and, consequently, the relative weight between bulk and surface. 

This is shown in figure \ref{aebulk_coulCV-SLY4}, which displays the behaviour of the bulk terms with the two prescriptions for the bulk density corresponding to consider the saturation density of infinite nuclear matter (curves labelled by ``sat''), or the central density of the ETF profile (curves labelled by ``cen''). For the latter choice, the bulk density depends on the mass number and it is systematically lower than the saturation density (see figure \ref{anprhocC}), leading to an increase of the nuclear bulk energy (left panel of figure \ref{aebulk_coulCV-SLY4}). The effect is smoothed with increasing isospin, and the impact is also negligible on the estimation of the Coulomb energy, as shown in the right panel of the same figure.  
 
\begin{figure}
\begin{center}
\includegraphics[scale=1.]{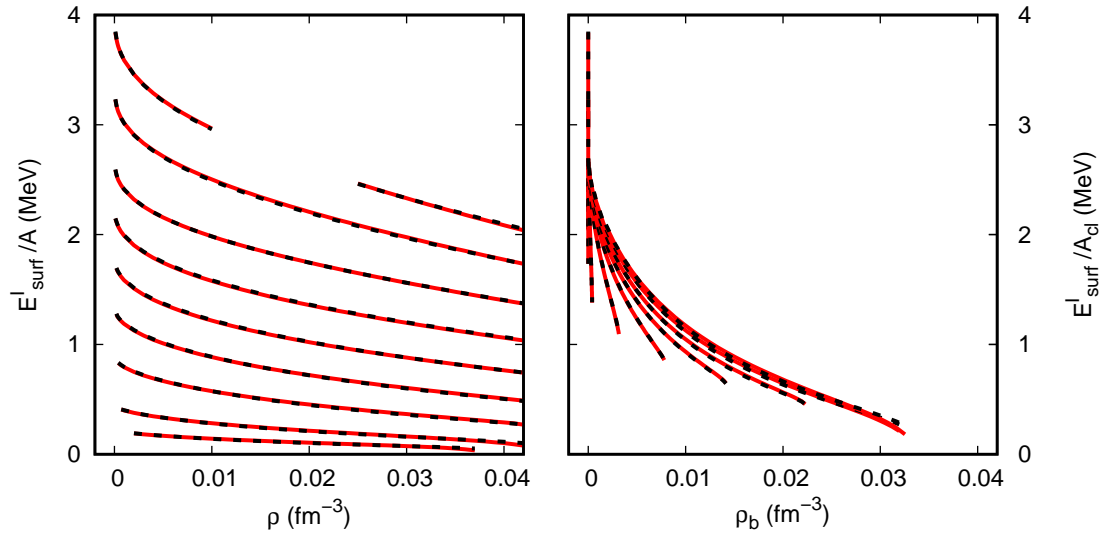}
\end{center}
\caption{Surface energy as a function of the total baryonic density (left) and of the external neutron gas density (right), for  
different isospin asymmetries from  $I=0.20$ (highest curve)  to $I=0.95$ (lowest curve), in steps of $\Delta I=0.1$.
 Solid lines: ETF calculations;  dashed lines: optimal fit using equation (\ref{eq:esurf_newton}). Saturation density was used (see text).
  Note that in the right panel the energy was divided by the mass number of the cluster.
}
\label{aesurfsatTC-SLY4}
\end{figure}

\begin{figure}
\begin{center}
\includegraphics[scale=1.]{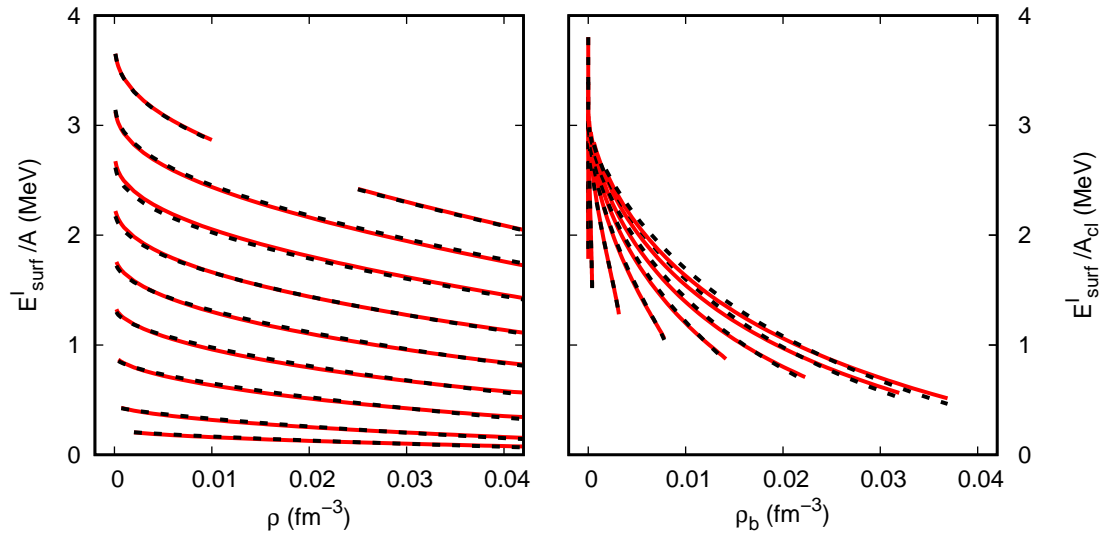}
\end{center}
\caption{Same as figure  \ref{aesurfsatTC-SLY4}, but the fit was done using  equation (\ref{eq:esurf_ravenhall}), 
and central density was used. 
}
\label{aesurfcenTC-SLY4}
\end{figure}

\subsubsection{Results: Nuclei in a gas}\label{res gas}
 
We now turn to the analysis of nuclei beyond the neutron dripline, signalled for each total proton number $Z=N_p$ by the change of sign of the neutron chemical potential. As already discussed in section \ref{res gas ETF}, for each isospin asymmetry value we varied the global density from $1.\times 10^{-4}\mathrm{fm}^{-3}$ to $4.2\times 10^{-2}\mathrm{fm}^{-3}$, such as to cover the typical density domain explored in neutron star crusts before the emergence of non-spherical pasta structures.

For each isospin asymmetry and global density,  the system of variational equations (\ref{eq:variational}) was solved, finding
not only the parameters of the FD profiles but also the background density and the
WS radius that provide the minimum energy, within the constraints given by the conservation laws equation  (\ref{N fixed}) and the charge neutrality.
The ETF surface energy was calculated by subtracting from the optimal ETF result the bulk energy given by equations (\ref{eq bulk gas}) and
(\ref{ex coulomb uniform 2}), with the two different prescriptions for the cluster density (saturation or central density).
The resulting surface energy was then fitted using  equation (\ref{eq:esurf_newton}) (for the saturation density choice) or equation (\ref{eq:esurf_ravenhall}) (for the central density choice).
  
The surface energy obtained in the ETF calculation and the corresponding optimal fits are shown in  figures \ref{aesurfsatTC-SLY4} and   \ref{aesurfcenTC-SLY4} as a function of the total baryonic density $\rho$, as well as the background neutron gas density $\rho_b$. 
In both figures, the missing points at the lowest asymmetry are due to the fact that  the nuclei obtained have isospin asymmetry below the stability according to the AME table, so they were not considered in the fit.
The corresponding values of the parameters are displayed, together with the $\chi^2$ of the fit and the number of iterations needed to achieve convergence, in tables  \ref{tesurfsatTC-SLY4}  and \ref{tesurfcenGTC-SLY4}, for saturation and central density, respectively. 
Similar to the results presented in section \ref{res vacuum}, we used the simpler equation (\ref{eq:curv}) for the curvature term appearing in equation (\ref{eq:esurf_newton}).

\begin{table}
\lineup
\caption{\label{tesurfsatTC-SLY4} Surface energy parameters fitted for nuclei from stability up to $I=0.95$ and quality of the corresponding fit for different choices of the $p$ parameter.  The last column gives the number of iterations needed to achieve convergence. The saturation density was employed.
  781 nuclei were considered. }
\begin{indented}
\item[]\begin{tabular}{@{}cccccc}
  \br
  $b_\mathrm{s}$ & $\sigma_0$(MeV$\cdot$fm$^ {-2}$) & $\sigma_\mathrm{0c}$(MeV$\cdot$fm$^ {-2}$) & $p$ & $\chi^2$ & iter \\
  \mr
\0\0  -1.92~ &  1.10159~ &  0.074341~ & 2.0~   &  116.438~ &  6 \\
\0\0  -1.23~ &  1.09381~ &  0.073179~ & 2.1~   &  100.496~ &  5 \\
\0\0  -0.35~ &  1.08605~ &  0.072057~ & 2.2~   & \085.638~ &  5 \\
\0\0\m   0.76~ &  1.07830~ &  0.070975~ & 2.3~ & \071.904~ &  5 \\
\0\0\m   2.15~ &  1.07059~ &  0.069934~ & 2.4~ & \059.328~ &  5 \\
\0\0\m   3.88~ &  1.06292~ &  0.068933~ & 2.5~ & \047.940~ &  5 \\
\0\0\m   6.01~ &  1.05530~ &  0.067974~ & 2.6~ & \037.763~ &  5 \\
\0\0\m   8.61~ &  1.04772~ &  0.067057~ & 2.7~ & \028.818~ &  4 \\
\0\m  11.77~ &  1.04020~ &  0.066180~ & 2.8~   & \021.119~ &  4 \\
\0\m  15.60~ &  1.03274~ &  0.065345~ & 2.9~   & \014.677~ &  4 \\
\0\m  20.22~ &  1.02535~ &  0.064550~ & 3.0~   & \0\09.498~ &  4 \\
\0\m  25.77~ &  1.01803~ &  0.063795~ & 3.1~   & \0\05.582~ &  4 \\
\0\m  32.41~ &  1.01079~ &  0.063080~ & 3.2~   & \0\02.929~ &  4 \\
\0\m  40.35~ &  1.00362~ &  0.062405~ & 3.3~   & \0\01.532~ &  4 \\
\0\m  49.82~ &  0.99654~ &  0.061768~ & 3.4~   & \0\01.382~ &  4 \\
\0\m  61.07~ &  0.98953~ &  0.061169~ & 3.5~   & \0\02.466~ &  4 \\
\0\m  74.44~ &  0.98262~ &  0.060608~ & 3.6~   & \0\04.770~ &  4 \\
\0\m  90.27~ &  0.97579~ &  0.060084~ & 3.7~   & \0\08.273~ &  5 \\
\0 109.02~ &  0.96906~ &  0.059596~ & 3.8~     & \012.957~ &  5 \\
\0 131.17~ &  0.96241~ &  0.059143~ & 3.9~     & \018.797~ &  5 \\
\0 157.31~ &  0.95586~ &  0.058725~ & 4.0~     & \025.769~ &  5 \\
\0 188.14~ &  0.94939~ &  0.058341~ & 4.1~     & \033.847~ &  5 \\
\0 224.45~ &  0.94302~ &  0.057991~ & 4.2~     & \043.002~ &  6 \\
\0 267.17~ &  0.93675~ &  0.057673~ & 4.3~     & \053.204~ &  6 \\
\0 317.41~ &  0.93057~ &  0.057387~ & 4.4~     & \064.423~ &  6 \\
\0 376.43~ &  0.92448~ &  0.057132~ & 4.5~     & \076.627~ &  6 \\
\0 445.71~ &  0.91848~ &  0.056908~ & 4.6~     & \089.785~ &  6 \\
\0 527.01~ &  0.91257~ &  0.056713~ & 4.7~     &  103.862~ &  6 \\
\0 622.33~ &  0.90676~ &  0.056548~ & 4.8~     &  118.827~ &  6 \\
\0 734.03~ &  0.90103~ &  0.056411~ & 4.9~     &  134.644~ &  7 \\
\0 864.87~ &  0.89539~ &  0.056302~ & 5.0~     &  151.282~ &  7 \\
\br
\end{tabular}
\end{indented}
\end{table}

\begin{table}
\lineup
  \caption{\label{tesurfcenGTC-SLY4} Surface energy parameters fitted for nuclei from stability up to $I=0.95$ and quality of the corresponding fit for different choices of the $p$ parameter.  The last column gives the number of iterations needed to achieve convergence. The central density was employed.
  786 nuclei were considered.}
\begin{indented}
\item[]\begin{tabular}{@{}cccccc}
  \br
  $b_\mathrm{s}$ & $\sigma_0$(MeV$\cdot$fm$^ {-2}$) & $\sigma_\mathrm{0c}$(MeV$\cdot$fm$^ {-2}$) & $p$ & $\chi^2$ & iter \\
  \mr
\0\0  -1.71~ &  1.07308~ &  0.030895~ &  2.0~   &  96.2390~ &   5 \\
\0\0  -1.05~ &  1.06712~ &  0.031116~ &  2.1~   &  87.5759~ &   5 \\
\0\0  -0.21~ &  1.06115~ &  0.031337~ &  2.2~   &  79.2577~ &   5 \\
\0\0\m   0.84~ &  1.05518~ &  0.031560~ &  2.3~ &  71.3068~ &   5 \\
\0\0\m   2.13~ &  1.04922~ &  0.031783~ &  2.4~ &  63.7442~ &   5 \\
\0\0\m   3.73~ &  1.04326~ &  0.032008~ &  2.5~ &  56.5897~ &   5 \\
\0\0\m   5.66~ &  1.03732~ &  0.032234~ &  2.6~ &  49.8617~ &   5 \\
\0\0\m   8.01~ &  1.03140~ &  0.032460~ &  2.7~ &  43.5773~ &   5 \\
\0\m  10.84~ &  1.02550~ &  0.032687~ &  2.8~   &  37.7520~ &   4 \\
\0\m  14.23~ &  1.01962~ &  0.032916~ &  2.9~   &  32.4001~ &   4 \\
\0\m  18.28~ &  1.01378~ &  0.033146~ &  3.0~   &  27.5343~ &   4 \\
\0\m  23.10~ &  1.00798~ &  0.033377~ &  3.1~   &  23.1661~ &   4 \\
\0\m  28.83~ &  1.00221~ &  0.033609~ &  3.2~   &  19.3052~ &   4 \\
\0\m  35.61~ &  0.99649~ &  0.033843~ &  3.3~   &  15.9603~ &   4 \\
\0\m  43.63~ &  0.99081~ &  0.034078~ &  3.4~   &  13.1384~ &   4 \\
\0\m  53.08~ &  0.98517~ &  0.034315~ &  3.5~   &  10.8453~ &   4 \\
\0\m  64.20~ &  0.97959~ &  0.034553~ &  3.6~   & \09.0854~ &   4 \\
\0\m  77.27~ &  0.97406~ &  0.034794~ &  3.7~   & \07.8619~ &   4 \\
\0\m  92.61~ &  0.96859~ &  0.035037~ &  3.8~   & \07.1766~ &   4 \\
\m 110.58~ &  0.96317~ &  0.035282~ &  3.9~     & \07.0302~ &   4 \\
\m 131.61~ &  0.95781~ &  0.035530~ &  4.0~     & \07.4223~ &   4 \\
\m 156.19~ &  0.95251~ &  0.035781~ &  4.1~     & \08.3513~ &   4 \\
\m 184.90~ &  0.94727~ &  0.036034~ &  4.2~     & \09.8146~ &   4 \\
\m 218.39~ &  0.94209~ &  0.036290~ &  4.3~     &  11.8088~ &   4 \\
\m 257.43~ &  0.93697~ &  0.036550~ &  4.4~     &  14.3292~ &   5 \\
\m 302.91~ &  0.93192~ &  0.036813~ &  4.5~     &  17.3705~ &   5 \\
\m 355.84~ &  0.92693~ &  0.037079~ &  4.6~     &  20.9266~ &   5 \\
\m 417.40~ &  0.92200~ &  0.037350~ &  4.7~     &  24.9906~ &   5 \\
\m 488.97~ &  0.91714~ &  0.037623~ &  4.8~     &  29.5549~ &   5 \\
\m 572.12~ &  0.91235~ &  0.037901~ &  4.9~     &  34.6112~ &   5 \\
\m 668.67~ &  0.90762~ &  0.038183~ &  5.0~     &  40.1508~ &   5 \\
\br
\end{tabular}
\end{indented}
\end{table}

In agreement with the results of section \ref{res vacuum}, we can see a clear decrease of the surface energy with the isospin, observed with both prescriptions for the surface energy.  For the  largest WS volumes, corresponding to $\rho\to 0$, we find back the results of section \ref{res vacuum} for stable nuclei. In this limit, the highest surface energies per nucleon are associated to the lighter nuclei. The decreasing behaviour of the surface energy with the density can be understood as an effect of the increasing importance of the background density, which smooths the interface between the nucleus and its environment and thus reduces the surface tension. Remarkably, this complex behaviour can be very well reproduced with a parametrization of the surface energy that only depends on the proton fraction of the cluster, and  does not depend on the external gas.  

Comparing figures \ref{aesurfsatTC-SLY4} and   \ref{aesurfcenTC-SLY4},
we can observe that both prescriptions to define the surface energy lead to excellent fits. This very interesting result means that, even in the free neutron regime, the explicit inclusion of different radius parameters for the proton and neutron density profiles, and the associated presence of skin neutrons, is not needed to have a precise and quantitative description of the isospin dependence of the surface tension: the simple prescription given by equations (\ref{eq:esurf_newton}), (\ref{eq sigma surf}) and (\ref{eq:curv}) for the surface energy, together with the estimation of the central equilibrium cluster density from
the isospin dependent saturation density of infinite nuclear matter equation (\ref{eq_asym_rho0}), are sufficient to correctly describe the surface energy and its isospin dependence up to almost pure neutron matter.  
A closer inspection of the quality of the fit
in tables \ref{tesurfsatTC-SLY4}  and \ref{tesurfcenGTC-SLY4}, surprisingly reveals that the fit is even better when we use the simpler prescription that ignores finite size effects in the bulk, the presence of neutrons in the skin, and effectively includes the existence of different radii into the isospin dependence of the surface energy.

The other interesting observation that we can get from tables \ref{tesurfsatTC-SLY4}  and \ref{tesurfcenGTC-SLY4} is that the $p$ parameter entering  equation (\ref{eq:esurf_newton}) is very important to get a good quality fit. This is at variance with the results of section  \ref{res vacuum}, where we saw that this parameter can be arbitrarily fixed (for instance to the $p=3$ value used in \cite{LS1991}) without affecting the quality of the fit. The present finding confirms the results of \cite{Carreau2019a,*Carreau2019b}, which showed that a careful optimization of the $p$ parameter is needed to describe highly asymmetric stellar matter close to the neutron star crust-core transition. In agreement with that work, optimal values of the $p$ parameter are found in the range $p\approx 3-4$ for the Sly4 interaction.

Concerning the value of the surface energy,  comparing figures \ref{aesurfsatTC-SLY4} and   \ref{aesurfcenTC-SLY4},
we can see that very close results are obtained in the two prescriptions. 
In the case of the central density fit (figure  \ref{aesurfcenTC-SLY4} and table \ref{tesurfcenGTC-SLY4}), we recall that the proton and neutron radii are not the same and a number $N_\mathrm{surf}$ of extra neutrons on the surface are considered in order to respect the particle number conservation, see  equation (\ref{n numb skin}). These extra neutrons modify the expression of the interface energy by adding an extra term $\mu_n N_\mathrm{surf}$, see equation (\ref{eq:esurf_ravenhall}). Since the total ETF energy is the same whatever the splitting between surface and bulk, the similar values of $E^I$ obtained with the two prescriptions means that the contribution of skin neutrons  when  equation (\ref{eq:esurf_ravenhall}) is used, is effectively accounted for in the bulk terms if equation (\ref{eq:esurf_newton}) is assumed.
  
 This observation can be understood as follows. At a given $(\rho,I)$ condition and for a given definition of the bulk density (central or saturation), equation (\ref{n numb skin}) indicates that larger cluster radii are obtained if we put $N_\mathrm{surf}=0$, that is, the nucleons in the interface are attributed to the cluster. This leads to a larger cluster size $A_{cl}$, as it can be seen from the fact that lower surface energies are obtained 
in the right panel of figure  \ref{aesurfsatTC-SLY4}, when the energies are normalized to the cluster size. The inclusion of surface neutrons in the definition of the cluster modifies also the cluster proton fraction $y_p$ and, consequently, the bulk energy.

\begin{figure}
\begin{center}
\includegraphics[scale=1.]{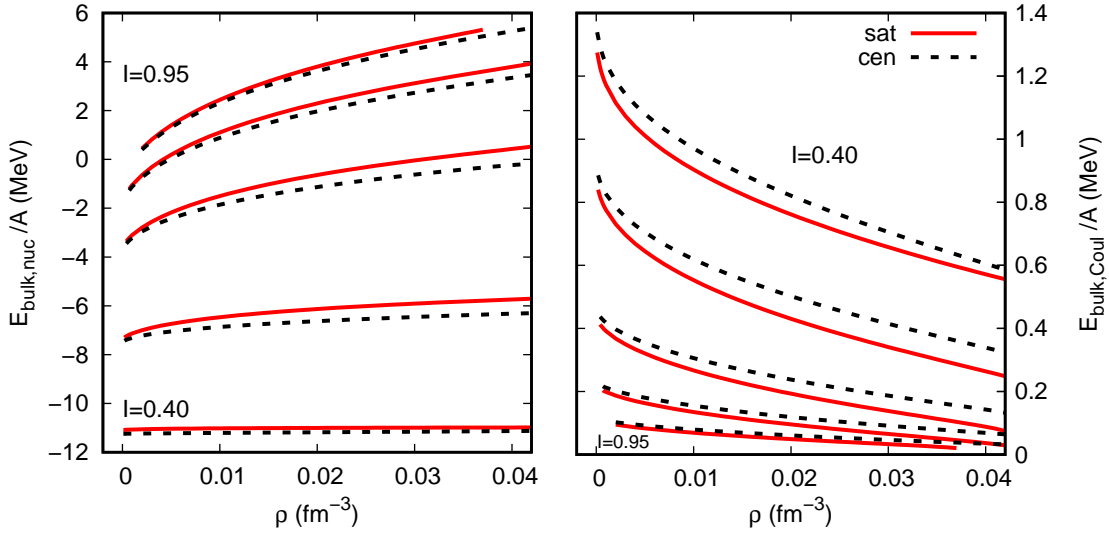}
\end{center}
\caption{Bulk nuclear (left part) and Coulomb (right part) energies per nucleon for nuclei in a gas as a function of the total density $\rho$, for isospin $I=0.40;~0.60;~0.80;~0.90$ and $0.95$. The behaviour with $I$ is monotonic. $A$ is the mass number of the whole WS cell. The bulk energy includes the energy from the gas, equation (\ref{eq bulk gas}).
}
\label{aebulk_coul-SLY4}
\end{figure}

This difference in the bulk terms is clearly seen in figure \ref{aebulk_coul-SLY4}, which displays the bulk terms in different  $(\rho,I)$ conditions, for the two prescriptions.
When the skin energetics is included in the bulk part (curve labelled ``sat''), more asymmetric clusters are obtained and, correspondingly, the nuclear binding is less important with respect to the ``cen'' choice, which considers these neutrons as part of the interface (left part of  figure \ref{aebulk_coul-SLY4}), even if this is partly compensated by the reduced Coulomb energy due to the increased cluster radius  (right part of  figure \ref{aebulk_coul-SLY4}).
Another source of difference between the two prescriptions lies in the definition itself of the bulk density. 
As already seen in figure \ref{arhocq-SLY4} above,
the saturation density is  different from the central density.  However, this  difference  is relatively small, and it produces a  negligible effect with respect to the one we have just discussed, due to the different cluster size obtained with the two prescriptions.

\begin{figure}
\begin{center}
\includegraphics[scale=1.]{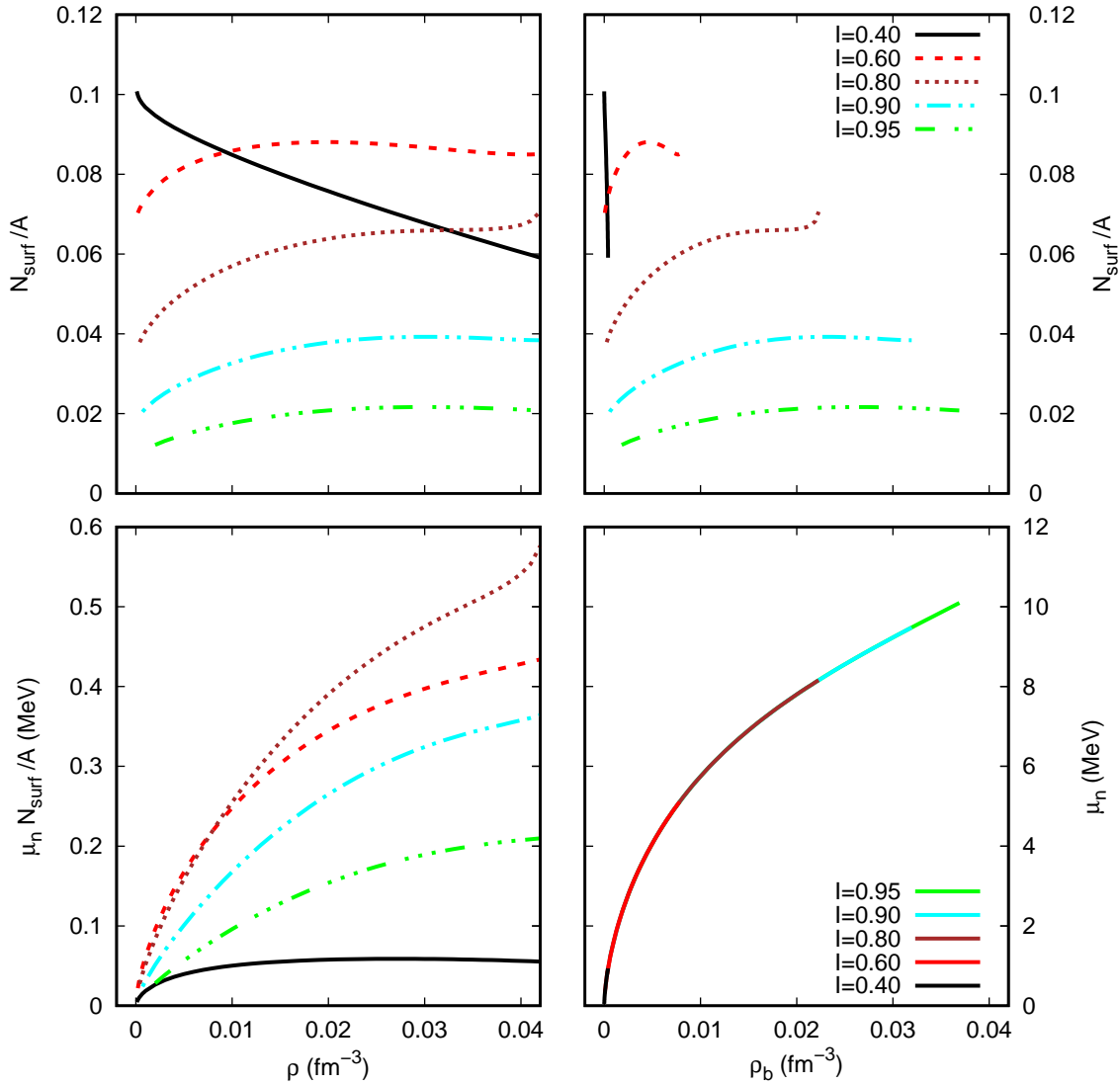}
\end{center}
\caption{Upper part: number of surface neutrons $N_\mathrm{surf}$ divided by the total cell baryon number for different values of isospin as a function of the total density (left), and as a function of the background density (right).
Lower right  part: neutron chemical potential $\mu_n$, for the same values of isospin as the upper part,
as a function of the background density.  The lower left part gives the product of the two quantities.
The central density was used (see text).
}
\label{amunNsurfcenC-SLY4}
\end{figure}
More details on the properties of the interface neutrons can be learnt from
figure \ref{amunNsurfcenC-SLY4}, which gives the behaviour of the neutron chemical potential and skin neutrons as a function of the total density $\rho$ and background density $\rho_{b}$. 
We recall that, in the free neutron regime, because of chemical equilibrium in the WS cell, $\mu_n$ can be identified with the chemical potential of the background gas, equation (\ref{eq:mugas}). We can see  from figure  \ref{amunNsurfcenC-SLY4} (lower panel) that $\mu_n$  monotonically increases with the isospin and is always positive for $I\ge 0.4$, implying that the contribution of the skin term $\mu_n N_\mathrm{surf}$ (lower left)
 is always positive too and its contribution is never negligible. 

Concerning the number of neutrons in the skin, at variance with the vacuum results shown in  figure  \ref{auNsurfcenPCV-SLY4}, no monotonic behaviour of $N_\mathrm{surf}$ is observed as a function of the density and isospin. This is understood from the complex behaviour of the cluster mass, which increases with the density and decreases with the isospin, and of the cluster asymmetry, which does not coincide with the global isospin asymmetry once the dripline is reached.  We can, first of all, notice a qualitatively different behaviour of the $I=0.4$ calculation, which is close to the dripline, and the ones corresponding to increasing neutron excess. For $I=0.4$, the decrease with the density of the fraction of skin neutrons is due to the increasing mass of the cluster, as we have observed in figure \ref{auNsurfcenPCV-SLY4} that larger nuclei have comparatively less skin neutrons. The mass increases with the density at fixed $I$ because only the most stable isotope for the given $(\rho, I)$ condition is obtained in the variational calculation, and the stability line is displaced with respect to the vacuum results due to the increased electron screening and, therefore, reduced Coulomb interaction.

Well above the dripline, at a fixed value of $I>0.4$, the fraction of surface neutrons increases with the total as well as background densities, and tends to saturate at high density. This approximate proportionality with the total number of particles explains why the complex behaviour  of nucleons in the interface can be recast in terms of a modified bulk term.  

\begin{figure}[h]
\begin{center}
\includegraphics[scale=1.]{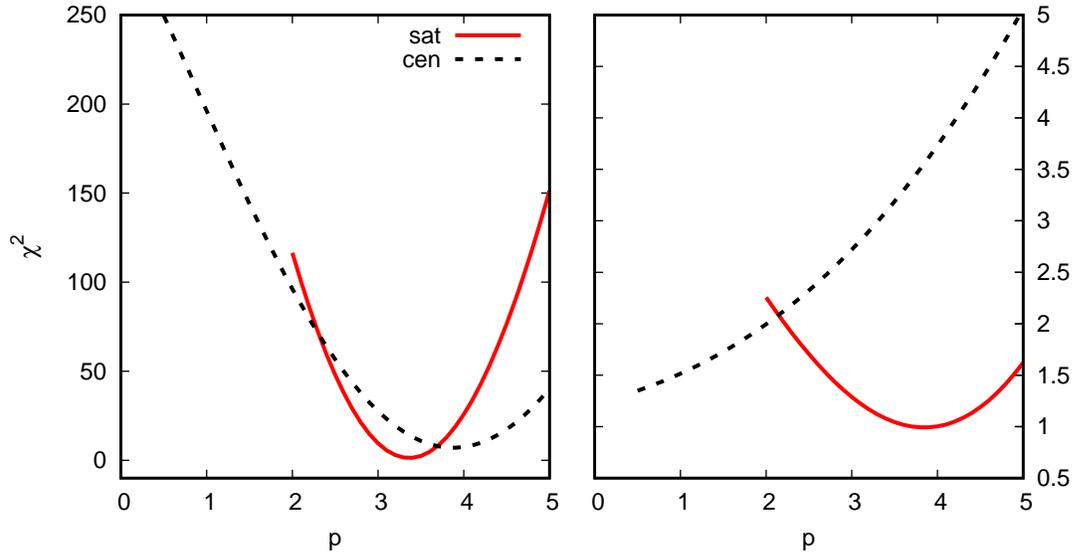}
\end{center}
\caption{Quality of the fit of the surface energy as measured by $\chi^2$ as a function of the parameter $p$. 
Full lines: equation (\ref{eq:esurf_newton}) and saturation density; dashed lines: equation (\ref{eq:esurf_ravenhall}) and central density.
Left panel: fit on all nuclei from stability up to $I=0.95$. Right side: fit up to the neutron dripline (nuclei in the vacuum).  
The results were taken from tables \ref{tpBC-sat}-\ref{tesurfcenGTC-SLY4}.
See text for more details.  
}
\label{achi2-SLY4}
\end{figure}

Figure \ref{achi2-SLY4} gives the quality of the fit and the final determination of the best $p$ parameter for the different techniques. 
The left panel of this figure shows the value of the optimal $\chi^2$ as a function of the $p$ parameter, when all other parameters entering the surface energy expressions, equations (\ref{eq:esurf_newton}) and (\ref{eq:esurf_ravenhall}), are optimized on the whole set of calculations including all nuclei from the stability line up to $I=0.95$.
We can see that both prescriptions lead to fits of comparable quality and allow a clear determination of the $p$ parameter that governs the behaviour of the surface tension at extreme isospin values. However, this value depends of the expression employed (skin neutrons included or not in the definition of the cluster), and on the prescription employed to fix the density of the bulk, namely the theoretical expression of the saturation density of infinite nuclear matter, or the variationally calculated central density of the nucleus. In neither case the value $p=3$ which has been widely used in the literature \cite{Ravenhall1983,LS1991,Lorenz1993,Lim2018,*Lim2019a,*Lim2019b} was obtained as optimal value.
 
To obtain those fits, we used the complete set of binding energy values, including nuclei well above the dripline whose masses can only be accessed through theoretical calculations in the Wigner-Seitz cell. 
If the surface energy is only optimized on binding energy of bound nuclei, one can wonder whether a realistic
 extrapolation to extreme isospin conditions, when neutrons are emitted in the continuum, is possible. This is shown in the right panel of figure \ref{achi2-SLY4}, which displays the quality of the fit when only nuclei up to the dripline are included in the fit.
We can see that the global quality of the fit is definitely better, but if we employ the full expression equation (\ref{eq:esurf_ravenhall}), which accounts for nucleons in the skin (dashed line),  and try to determine the surface energy using only the information of the mass of terrestrial nuclei, we cannot determine an optimal $p$ parameter.
With the same limited information on bound nuclei only, this is however possible if  the simpler expression equation (\ref{eq:esurf_newton}) 
 is employed, and the bulk density is assumed to be given by the saturation density (full line in the same panel).

\begin{table}
\lineup
  \caption{\label{tab:param} Optimal parameters for the different hypotheses. See text.}
\begin{indented}
\item[]\begin{tabular}{@{}ccccccc}
  \br
 & $condition$ & $b_\mathrm{s}$ & $\sigma_0$ & $\sigma_\mathrm{0c}$ & $p$  &$\chi^2$\\
  \mr
(A) & vacuum, no skin~ &  102.745~ &  0.95720~ &  0.073773~  & 3.8~ &  0.9930 \\
(B) & vacuum, skin~    &  \0\0-2.431~  &  0.93252~ &  0.058036~  & 0.5~ &  1.3492 \\
(A') & $I<0.95$, no skin ~ &  \049.82~  &  0.99654~ &  0.061768~  &  3.4~  &  1.382 \\
(B') & $I<0.95$, skin~ &  110.58~  &  0.96317~  &  0.035282~  &  3.9~  &  7.0302  \\
\br
\end{tabular}
\end{indented}
\end{table} 

\begin{figure}
\begin{center}
\includegraphics[scale=1.]{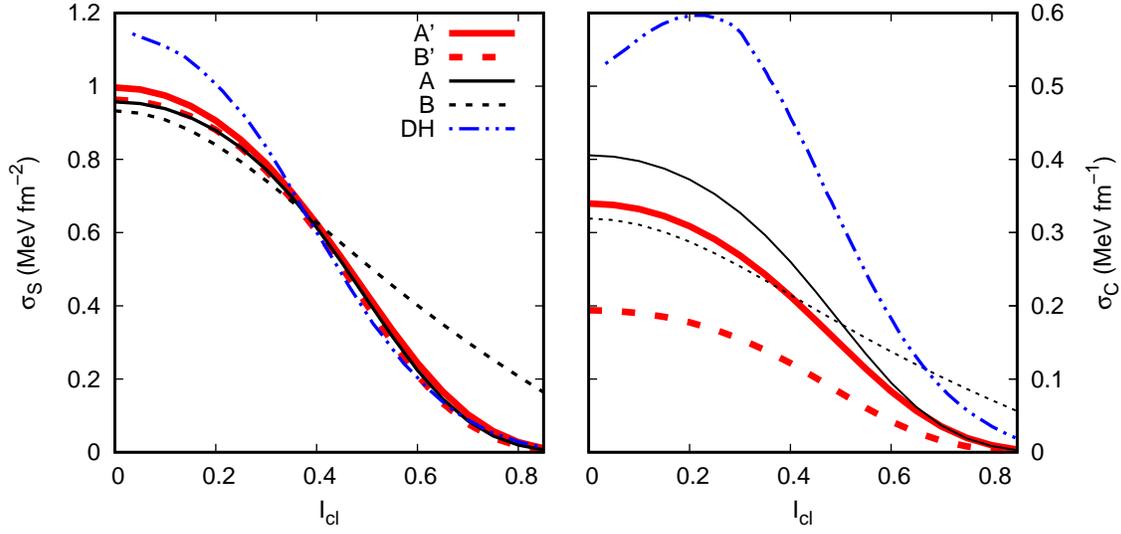}
\end{center}
\caption{
Analytical representation of the surface tension (left panel) and curvature (right panel)  with the Sly4 interaction vs $I_{cl}$,  for the different conditions given in table \ref{tab:param}. Thin black lines: fit from nuclei in the vacuum. Thick red lines: fit from all nuclei up to $I=0.95$. Solid lines: saturation density and no neutrons at the interface. Dashed lines: central density and neutrons at the interface. The blue dash-dotted lines give the results of the DH model \cite{Douchin2001}.
}
\label{aDH-SLY4}
\end{figure}

To summarize our results, figure \ref{aDH-SLY4}
displays the behaviour of the analytical expression for the surface tension (left part) and curvature parameter (right part),  for the different hypotheses and conditions. The corresponding optimal parameters, which are the ones corresponding to  the minimum 
$\chi^2$,  are given in  table \ref{tab:param}. 

Comparing the thick full (conditions (A') in table \ref{tab:param}) and the thick dashed (condition (B') in table \ref{tab:param})  lines, we can appreciate the effect of the two different prescriptions for the surface energy, 
 equations (\ref{eq:esurf_newton}) and (\ref{eq:esurf_ravenhall}).

We can see that the two surface tensions are well compatible, while the curvature term obtained using the saturation density is systematically higher than the one using the central density and including explicitly the contribution of the skin neutrons. 
As a consequence, the latter prescription leads to a global surface energy that is slightly lower for the lighter nuclei, for which the curvature term cannot be neglected. 
As we have already observed, this systematic difference, which is only reduced at extreme isospin values, is due to the different decomposition of the total energy into bulk and surface.  
However, if the bulk energy is consistently treated within each prescription, the two descriptions lead to equivalently good representations of the nuclear energetics. 

If we now compare the thick lines with the thin lines, we can appreciate the quality of the extrapolation towards neutron matter of a surface energy optimized on nuclei below the driplines, such as the ones that can be produced in laboratory experiments. 

In the case of the saturation density and equation (\ref{eq:esurf_newton}) (full lines, conditions (A) and (A')),
we can see in the left panel that, in spite of the difference in the parameters, the surface tension optimized on bound nuclei is perfectly compatible with the one optimized on calculations including a neutron gas. Looking at the right panel, we can see 
that a difference appears in the curvature term, especially for the lower values of isospin, below $I=0.4$: the optimization to the whole set of nuclei including a background gas (full thick lines) leads to an underestimation of the curvature term for the nuclei below the dripline. Conversely, we can say that the much simpler fit on nuclei in the vacuum (full thin lines), which, as we saw in section \ref{res vacuum}, reproduces the ETF energy with remarkable accuracy, can be reasonably extrapolated to describe nuclei beyond the dripline, with only a slight overestimation of the curvature term.

The same is not true for the central density choice and  equation (\ref{eq:esurf_ravenhall}) (dashed lines), where we can see that an optimization on bound nuclei (thin lines)  leads to a very poor extrapolation towards neutron matter.  In this case, the non-realistic extrapolation concerns both the curvature and the surface tension, meaning that it will affect nuclei of all sizes, including pasta structures in the innermost part of the inner crust. This problem cannot be satisfactorily solved by using the optimization to the whole set of nuclei (dashed thick lines): in that latter case, realistic results for extreme isospin values are obtained at the price of an important deviation of the curvature term  at small isospin. We can therefore conclude that the prescription given by equation (\ref{eq:esurf_newton}), using a fit that is limited to bound nuclei within the driplines (condition (A), full thin lines), is not only simpler, but also more realistic.

Finally, the dash-dotted lines in figure \ref{aDH-SLY4} give the surface tension and curvature parameter of the popular DH model \cite{Douchin2001}, which is based on the same Sly4 interaction as in the present study. In the DH paper, similar to the previous seminal  LLPR model \cite{lattimer1985}, the surface tension was extracted from slab calculations, and the curvature term was computed in perturbation. We can see that this perturbative procedure leads to a surface energy that is systematically higher than a direct fit on finite nuclei and Wigner-Seitz cells.

However,
there is an almost  perfect agreement between our results with the fully analytical surface model, equation (\ref{eq:esurf_newton}) (full lines), and the DH results, as far as the surface tension is concerned, and isospin values beyond drip are considered. 
We therefore consider that our results are fully compatible with the DH analysis. Indeed,  
the DH model is conceived to be applied to the inner crust, where $I>0.4$ and the clusters are so massive that the curvature term plays a very small role.

\begin{figure}
\begin{center}
\includegraphics[scale=1.]{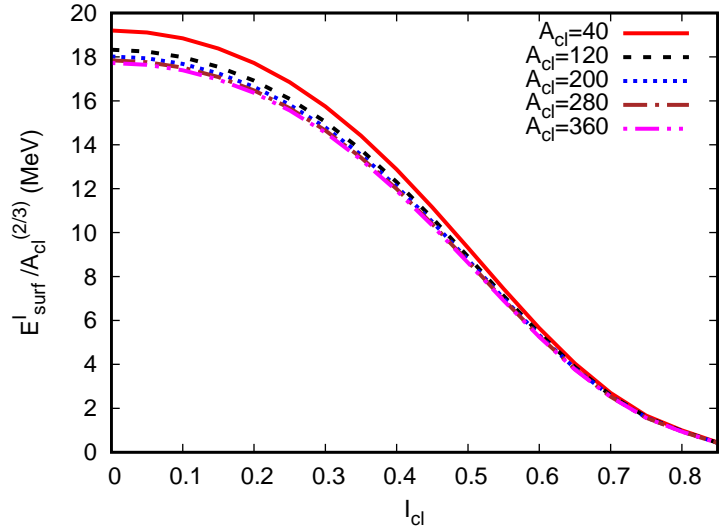}
\end{center}
\caption{Final result for the analytical representation of the surface tension $E_\mathrm{surf}^I/A_{cl}^{2/3}$ of spherical nuclei with the Sly4 interaction vs $I_{cl}$, for different mass numbers from $A_{cl}=40$ (top line) to $A_{cl}=360$ (bottom line), in steps of $\Delta A=80$. The corresponding parameters are given on the third line of table \ref{tab:param}.
}
\label{anuclei-SLY4}
\end{figure}

The validity of this last  statement can be appreciated from figure \ref{anuclei-SLY4}, which 
shows our final result for the surface energy $E^I_\mathrm{surf}/A_{cl}^{2/3}$ for different mass numbers, 
from the analytical expression equation (\ref{eq:esurf_newton}).
The saturation density was employed, and the parameter values are such as to minimize the $\chi^2$.
We can observe that the mass dependence due to the curvature term is important for light nuclei close to stability, but it becomes less and less important as the nuclei become more massive, and the isospin asymmetry increases.

For this result, as for the rest of our analysis, we employed the Sly4 functional. 
The absolute value and  behaviour of the surface tension  obviously depends on the functional, and a detailed study of its model dependence  is left for future work.

\section{Conclusions and outlooks}\label{sec conclusions}

In this paper we presented extended Thomas-Fermi calculations at second order in $\hbar$  in spherical symmetry, with the purpose of studying and parametrizing the surface tension of extremely neutron rich nuclei well beyond the dripline, for applications in the sub-saturation regime of the equation of state. These nuclei are explored in different astrophysical environments, and modifications of their properties are expected in the dense medium constituted by their continuum states. In most equations of state, these in-medium modifications are modelled in the excluded volume approximation, but modifications of the surface tension due to the external nucleon gas might also be expected. 

In this work, we neglected the presence of a proton gas, which limits the application to moderate temperatures. We showed that the simple Ravenhall {\it et al}. \cite{Ravenhall1983} expression can reproduce with remarkable accuracy the surface energy of  nuclei in the large mass ($A=40-650$)  and isospin ($I=0.20-0.95$) ranges. The reduction of the surface tension due to the external neutron gas is seen to solely depend on the proton fraction of the nucleus, provided a parameter $p$ governing the extreme isospin dependence is introduced and optimized on calculations extended at least up to the dripline.

Only spherical nuclei were considered in this work. However, we showed that the importance of the curvature term strongly decreases not only with the mass, but also with the isospin of the nucleus. At very large values of isospin $I\approx 0.5$, this geometry-dependent term can be neglected and the resulting surface tension can be applied to arbitrary geometries to describe deformed pasta phases.
  
To clarify the role of the nuclear skin on the surface tension, two different expressions for the surface energy were analyzed. The first one introduces two different effective radii for the neutron and proton distributions, leading to a number of neutrons at the nucleus skin that increases with the mass and isospin of the cluster. In this picture,  the energetic contribution of the skin nucleons is explicitly accounted for, and it is seen to depend in a complex way on both the variables of the cluster and the variables of the gas. 
Surprisingly, this more elaborate modelling does not lead to a more precise reproduction of the total ETF variational energy, and a very precise mass formula can be obtained up to almost pure neutron matter with a simple decomposition of the total energy into standard bulk and surface terms.

A quantitative comparison with the surface energy of the popular DH model for the neutron star inner crust showed a perfect compatibility in the isospin and mass regimes explored in the inner crust, but sizeable differences are observed for small and relatively symmetric nuclei.  Such nuclei are mostly produced in supernova matter, that is, at finite temperature and relatively higher proton fractions than the ones explored in full beta equilibrum. For all these applications, we believe that the simple analytical formula proposed in the present work can give a very realistic prescription for the surface energy.

In this paper, we concentrated on the popular Sly4 functional. However, our formalism based on a  meta-modelling of the equation of state can be extended to different functionals. A detailed study of the model dependence of the surface tension will be presented in a forthcoming paper.

\ack

This work was partially conducted during a scholarship supported by the International Cooperation Program
CAPES/COFECUB Fondation, agreement Ph853-15, at the University of Caen Normandy. Financed by CAPES – Brazilian Federal Agency
for Support and Evaluation of Graduate Education within the Ministry of Education of Brazil.
Furtado U J is thankful to Menezes D P for her support and discussions.

\appendix

\section{Variational derivation below neutron drip}\label{appA}

In this Appendix, we give the detailed expression of the coupled variational equations (\ref{eq zero}) that have to be solved to obtain the optimal density profiles and the associated energy below the neutron dripline. 
We recall that the input parameters are the mass $A$ and proton $N_p=Z$ numbers of the nucleus, and the variational variables to be optimized are the profile radii, diffusivities, and central densities: $R_n,~R_p,~a_n,~a_p,~\rho_{cn}$ and $\rho_{cp}$.
Out of these variables, two of them should be fixed by imposing particle number conservation. 
We choose the radii $R_n,~R_p$.

We can use equation  (\ref{N fixed 2}) and write ($z_{qi}=a_n,~a_p,~\rho_{cn}$ and $\rho_{cp}$):

\begin{equation}
\frac{\partial g}{\partial z_{qi}}= \left. \left (\frac{\partial g}{\partial R_q} \right )\right|_{z_{qi}=\mathrm{const}}
\frac{\partial R_q}{\partial {z_{qi}}}
+ \left. \left ( \frac{\partial g}{\partial {z_{qi}}} \right )\right|_{R_q=\mathrm{const}}.
\end{equation}

For simplicity, to write the derivatives in the following, we will consider the $R_q$'s as
independent variables and then come back to the above equation.
We denote the zero $\hbar$ order kinetic energy by $\hbar^2(\tau_{0n}/2m^*_n+\tau_{0p}/2m^*_p)\equiv t^{\mathrm{FG}^*}$.
This term does not contain density derivatives. Making use of the
chain rule, we can write ($y_{qi}=R_n,~R_p,~a_n,~a_p,~\rho_{cn}$ and $\rho_{cp}$):

\begin{equation}
\frac{\partial}{\partial y_{qi}} (t^{\mathrm{FG}^*})= \frac{\partial}{\partial \rho_q} (t^{\mathrm{FG}^*})
\frac{\partial \rho_q}{\partial y_{qi}};
\end{equation}

\beqa
\frac{\partial}{\partial \rho_n} (t^{\mathrm{FG}^*})= \frac{3 \hbar^2}{10} (3 \pi^2)^{2/3}
\left( \frac{5}{3} \frac{\rho_n^{2/3}}{m^*_n} +
\kappa_+ \frac{\rho_n^{5/3}}{m} + \kappa_- \frac{\rho_p^{5/3}}{m} \right);
\eeqa

\beqa
\frac{\partial}{\partial \rho_p} (t^{\mathrm{FG}^*})= \frac{3 \hbar^2}{10} (3 \pi^2)^{2/3}
\left( \frac{5}{3} \frac{\rho_p^{2/3}}{m^*_p} +
\kappa_+ \frac{\rho_p^{5/3}}{m} + \kappa_- \frac{\rho_n^{5/3}}{m} \right),
\eeqa
where
\beqa
\kappa_+= \frac{1}{\rho_\mathrm{sat}} (\kappa_\mathrm{sat}+\kappa_\mathrm{sym}); \\
\kappa_-= \frac{1}{\rho_\mathrm{sat}} (\kappa_\mathrm{sat}-\kappa_\mathrm{sym}).
\eeqa

For the potential energy part, the case is similar:

\begin{equation}
\frac{\partial v}{\partial y_{qi}} = \frac{\partial v}{\partial \rho_q} 
 \frac{\partial \rho_q}{\partial y_{qi}};
\end{equation}

\beqaa 
\frac{\partial v}{\partial \rho_n} =& \frac{v}{\rho} \left(1-\frac{b\rho}{\rho_\mathrm{sat}} \right) +
\rho \sum_{\alpha=0}^N \frac{1}{\alpha !} \cdot \nonumber \\
&\left \{ 4v_\alpha^\mathrm{iv} \delta \frac{\rho_p}{\rho^2} x^\alpha u_\alpha^N +
\frac{1}{\rho_\mathrm{sat}}(v_\alpha^\mathrm{is} + v_\alpha^\mathrm{iv} \delta^2)
\left[\frac{\alpha}{3}x^{\alpha-1} u_\alpha^N \right. \right. \nonumber \\
&+ \left. \left. x^\alpha(N+1-\alpha)(-3x)^{N-\alpha} \exp(-b\rho / \rho_\mathrm{sat}) + b x^\alpha \right] \right \};
\eeqaa 

\beqaa 
\frac{\partial v}{\partial \rho_p} =& \frac{v}{\rho} \left(1-\frac{b\rho}{\rho_\mathrm{sat}} \right) +
\rho \sum_{\alpha=0}^N \frac{1}{\alpha !} \cdot \nonumber \\
&\left \{-4v_\alpha^\mathrm{iv} \delta \frac{\rho_n}{\rho^2} x^\alpha u_\alpha^N +
\frac{1}{\rho_\mathrm{sat}}(v_\alpha^\mathrm{is} + v_\alpha^\mathrm{iv} \delta^2)
\left[\frac{\alpha}{3}x^{\alpha-1} u_\alpha^N \right. \right. \nonumber \\
&+ \left. \left. x^\alpha(N+1-\alpha)(-3x)^{N-\alpha} \exp(-b\rho / \rho_\mathrm{sat}) + b x^\alpha \right] \right \}.
\eeqaa 

For the Coulomb part:

\beqaa 
\frac{\partial}{\partial y_{pi}} ({\mathcal H}_\mathrm{Coul})=&
\frac{e^2}{2} \frac{\partial \rho_p(r)}{\partial y_{pi}}
\left[ \int_0^r \rho_p(r') \frac{{r'}^2}{r} dr' + \int_r^\infty \rho_p(r') r' dr'\right] \nonumber \\
&+ \frac{e^2}{2} \rho_p(r) \left[ \int_0^r \frac{\partial \rho_p(r')}{\partial y_{pi}}
\frac{{r'}^2}{r} dr' + \int_r^\infty \frac{\partial \rho_p(r')}{\partial y_{pi}} r' dr'\right] \nonumber \\
&- \frac{e^2}{4\pi} \left(\frac{3\rho_p(r)}{\pi} \right)^{1/3} \frac{\partial \rho_p(r)}{\partial y_{pi}}.
\eeqaa 

For the local second $\hbar$ order kinetic term $\tau_{2q}^l$:

\beqaa 
\frac{\partial}{\partial y_{ni}}& \left(\sum_{q=n,p} \frac{\hbar^2}{2 m_q^*} \tau_{2q}^l \right)
=
\frac{\hbar^2}{2}\frac{1}{36} \left[\frac{\kappa_+}{m}
\frac{(\nabla \rho_n)^2}{\rho_n} +
\frac{\kappa_-}{m} \frac{(\nabla \rho_p)^2}{\rho_p} \right. \nonumber \\
&- \left. \frac{1}{m_n^*}\frac{(\nabla \rho_n)^2}{\rho_n^2} \right] \frac{\partial \rho_n}{\partial y_{ni}} +
\frac{\hbar^2}{2}\frac{1}{36}\frac{2}{m_n^*} \frac{\rho'_n}{\rho_n} \frac{\partial \rho'_n}{\partial y_{ni}}
\nonumber \\
&+ \frac{\hbar^2}{2}\frac{1}{3}\left[\frac{\kappa_+}{m} \nabla^2\rho_n+
\frac{\kappa_-}{m} \nabla^2\rho_p \right]
\frac{\partial \rho_n}{\partial y_{ni}}
+ \frac{\hbar^2}{2}\frac{1}{3}\frac{1}{m_n^*}\frac{\partial }{\partial y_{ni}}(\nabla^2\rho_n);
\eeqaa 

\beqaa 
\frac{\partial}{\partial y_{pi}}& \left(\sum_{q=n,p} \frac{\hbar^2}{2 m_q^*} \tau_{2q}^l \right)
=
\frac{\hbar^2}{2}\frac{1}{36} \left[\frac{\kappa_+}{m}
\frac{(\nabla \rho_p)^2}{\rho_p} +
\frac{\kappa_-}{m} \frac{(\nabla \rho_n)^2}{\rho_n} \right. \nonumber \\
&- \left. \frac{1}{m_p^*}\frac{(\nabla \rho_p)^2}{\rho_p^2} \right] \frac{\partial \rho_p}{\partial y_{pi}} +
\frac{\hbar^2}{2}\frac{1}{36}\frac{2}{m_p^*} \frac{\rho'_p}{\rho_p} \frac{\partial \rho'_p}{\partial y_{pi}} \nonumber \\
&+ \frac{\hbar^2}{2}\frac{1}{3}\left[\frac{\kappa_+}{m} \nabla^2\rho_p +
\frac{\kappa_-}{m} \nabla^2\rho_n \right]
\frac{\partial \rho_p}{\partial y_{pi}}
+ \frac{\hbar^2}{2}\frac{1}{3}\frac{1}{m_p^*}\frac{\partial }{\partial y_{pi}}(\nabla^2\rho_p),
\eeqaa 
where we have denoted
\begin{equation}
\rho'_q=\frac{\partial \rho_q}{\partial r}.
\end{equation}

For the non-local second $\hbar^2$ order kinetic term $\tau_{2q}^{nl}$:

\beqaa 
\frac{\partial}{\partial y_{ni}}& \left(\sum_{q=n,p} \frac{\hbar^2}{2 m_q^*} \tau_{2q}^{nl} \right)
=
\frac{\hbar^2}{12m} \left\{ \left[\nabla f_n + \left(\kappa_+\nabla\rho_n
+ \kappa_-\nabla\rho_p\right) \right]
\frac{\partial}{\partial y_{ni}}(\nabla\rho_n) \right. \nonumber \\
&+ \frac{\partial \rho_n}{\partial y_{ni}} \nabla^2 f_n +
(\rho_n \kappa_+ + \rho_p \kappa_-)\left[\frac{2}{r} \frac{\partial}{\partial y_{ni}}(\nabla\rho_n) +
\frac{\partial}{\partial y_{ni}}\left(\frac{\partial^2\rho_n}{\partial r^2} \right) \right]
\nonumber \\
&+ \frac{1}{2}\left(\frac{(\nabla f_n)^2}{f_n} -\kappa_+
\rho_n \frac{(\nabla f_n)^2}{f_n^2} -\kappa_-
\rho_p \frac{(\nabla f_p)^2}{f_p^2}\right) \frac{\partial \rho_n}{\partial y_{ni}}
\nonumber \\
&+ \left. \left(\kappa_+\rho_n\frac{\nabla f_n}{f_n} + \kappa_-\rho_p\frac{\nabla f_p}{f_p} \right)
\frac{\partial}{\partial y_{ni}}(\nabla\rho_n) \right\};
\eeqaa 

\beqaa 
\frac{\partial}{\partial y_{pi}}& \left(\sum_{q=n,p} \frac{\hbar^2}{2 m_q^*} \tau_{2q}^{nl} \right)
=
\frac{\hbar^2}{12m} \left\{ \left[\nabla f_p + \left(\kappa_+\nabla\rho_p
+ \kappa_-\nabla\rho_n\right) \right]
\frac{\partial}{\partial y_{pi}}(\nabla\rho_p) \right. \nonumber \\
&+ \frac{\partial \rho_p}{\partial y_{pi}} \nabla^2 f_p +
(\rho_p \kappa_+ + \rho_n \kappa_-) \left[\frac{2}{r} \frac{\partial}{\partial y_{pi}}(\nabla\rho_p) +
\frac{\partial}{\partial y_{pi}}\left(\frac{\partial^2\rho_p}{\partial r^2} \right) \right]
\nonumber \\
&+ \frac{1}{2}\left(\frac{(\nabla f_p)^2}{f_p} -\kappa_+
\rho_p \frac{(\nabla f_p)^2}{f_p^2} -\kappa_-
\rho_n \frac{(\nabla f_n)^2}{f_n^2}\right) \frac{\partial \rho_p}{\partial y_{pi}}
\nonumber \\
&+ \left. \left(\kappa_+\rho_p\frac{\nabla f_p}{f_p} + \kappa_-\rho_n\frac{\nabla f_n}{f_n} \right)
\frac{\partial}{\partial y_{pi}}(\nabla\rho_p) \right\}.
\eeqaa 

For the spin-orbit part:

\beqaa 
\frac{\partial}{\partial y_{ni}}({\mathcal H}_{\mathrm{SO}})=& -\frac{m}{\hbar^2}W_0^2 \left[
\left(\frac{1}{f_n}-\frac{\rho_n}{f_n^2}\kappa_+ \right)
\frac{\partial \rho_n}{\partial y_{ni}} \cdot \right. \nonumber \\
&\left( (\nabla \rho_n)^2 + \frac{(\nabla \rho_p)^2}{4} + \nabla \rho_n \nabla \rho_p \right) \nonumber \\
&\left. + \frac{\rho_n}{f_n} \left( 2\nabla \rho_n 
+ \nabla \rho_p \right) \frac{\partial}{\partial y_{ni}}(\nabla\rho_n) \right] \nonumber \\
&-\frac{m}{\hbar^2}W_0^2 \left[
-\frac{\rho_p}{f_p^2}\kappa_-\frac{\partial \rho_n}{\partial y_{ni}} \right.
\left( (\nabla \rho_p)^2 + \frac{(\nabla \rho_n)^2}{4} + \nabla \rho_n \nabla \rho_p \right) \nonumber \\
&\left. + \frac{\rho_p}{f_p} \left( \frac{\nabla \rho_n}{2}
+ \nabla \rho_p \right) \frac{\partial}{\partial y_{ni}}(\nabla\rho_n) \right];
\eeqaa 

\beqaa 
\frac{\partial}{\partial y_{pi}}({\mathcal H}_{\mathrm{SO}})=& -\frac{m}{\hbar^2}W_0^2 \left[
\left(\frac{1}{f_p}-\frac{\rho_p}{f_p^2}\kappa_+ \right)
\frac{\partial \rho_p}{\partial y_{pi}} \cdot \right. \nonumber \\
&\left( (\nabla \rho_p)^2 + \frac{(\nabla \rho_n)^2}{4} + \nabla \rho_p \nabla \rho_n \right) \nonumber \\
&\left. + \frac{\rho_p}{f_p} \left( 2\nabla \rho_p
+ \nabla \rho_n \right) \frac{\partial}{\partial y_{pi}}(\nabla\rho_p) \right] \nonumber \\
&-\frac{m}{\hbar^2}W_0^2 \left[
-\frac{\rho_n}{f_n^2}\kappa_- \frac{\partial \rho_p}{\partial y_{pi}} \right.
\left( (\nabla \rho_n)^2 + \frac{(\nabla \rho_p)^2}{4} + \nabla \rho_p \nabla \rho_n \right) \nonumber \\
&\left. + \frac{\rho_n}{f_n} \left( \frac{\nabla \rho_n}{2}
+ \nabla \rho_n \right) \frac{\partial}{\partial y_{pi}}(\nabla\rho_p) \right].
\eeqaa

Derivatives of the densities and its derivatives with respect to $y_{qi}$ are written as:

\beqa
\rho'_q=\frac{\partial \rho_q}{\partial r} = -\frac{\rho_{cq}}{a_q}
\frac{\exp[(r-R_q)/a_q]}{(1+ \exp[(r-R_q)/a_q])^2};
\eeqa

\beqa
\nabla^2\rho_q=\frac{2}{r}\rho'_q + \frac{\partial^2 \rho_q}{\partial r^2};
\eeqa

\beqa
\frac{\partial^2 \rho_q}{\partial r^2}= -\frac{\rho_{cq}}{a_q^2}\exp[(r-R_q)/a_q]
\frac{1-\exp[(r-R_q)/a_q]}{(1+ \exp[(r-R_q)/a_q])^3}.
\eeqa

\noindent Explicitating to the case  $y_{qi}=\rho_{cq}$ we have:
\beqa
\frac{\partial \rho_q}{\partial \rho_{cq}} = \frac{1}{1+ \exp[(r-R_q)/a_q]};
\eeqa

\beqa
\frac{\partial \rho'_q}{\partial \rho_{cq}} = -\frac{1}{a_q}
\frac{\exp[(r-R_q)/a_q]}{(1+ \exp[(r-R_q)/a_q])^2};
\eeqa

\beqa
\frac{\partial}{\partial \rho_{cq}} \left(\frac{\partial^2 \rho_q}{\partial r^2}\right)=
-\frac{\exp[(r-R_q)/a_q]}{a_q^2}
\frac{1-\exp[(r-R_q)/a_q]}{(1+ \exp[(r-R_q)/a_q])^3}.
\eeqa

\noindent Considering $y_{qi}=R_q$ we have:
\begin{equation}
\frac{\partial \rho_q}{\partial R_q} = \frac{\rho_{cq}}{a_q}
\frac{\exp[(r-R_q)/a_q]}{(1+ \exp[(r-R_q)/a_q])^2};
\end{equation}

\begin{equation}
\frac{\partial \rho'_q}{\partial R_q} = \frac{\rho_{cq}}{a_q^2}\exp[(r-R_q)/a_q]
\frac{1-\exp[(r-R_q)/a_q]}{(1+ \exp[(r-R_q)/a_q])^3};
\end{equation}

\beqaa 
\frac{\partial}{\partial R_q} \left(\frac{\partial^2 \rho_q}{\partial r^2}\right)=&
\frac{\rho_{cq}}{a_q^3}\exp[(r-R_q)/a_q] \cdot \nonumber \\
&\left[\frac{1-4\exp[(r-R_q)/a_q] + \rme^{2(r-R_q)/a_q}}{(1+ \exp[(r-R_q)/a_q])^4} \right];
\eeqaa 

\noindent And finally for $y_{qi}=a_q$ we have:
\begin{equation}
\frac{\partial \rho_q}{\partial a_q} = \frac{\rho_{cq}}{a_q^2}(r-R_q)
\frac{\exp[(r-R_q)/a_q]}{(1+ \exp[(r-R_q)/a_q])^2};
\end{equation}

\beqaa 
\frac{\partial \rho'_q}{\partial a_q} =& \frac{\rho_{cq}}{a_q^2}
\frac{\exp[(r-R_q)/a_q]}{(1+ \exp[(r-R_q)/a_q])^3} \cdot \nonumber \\
&\left[1+\exp[(r-R_q)/a_q] + \frac{r-R_q}{a_q}(1-\exp[(r-R_q)/a_q]) \right];
\eeqaa 

\beqaa 
\frac{\partial}{\partial a_q} \left(\frac{\partial^2 \rho_q}{\partial r^2}\right)=&
\frac{\rho_{cq}}{a_q^4}
\frac{\exp[(r-R_q)/a_q]}{(1+ \exp[(r-R_q)/a_q])^4} \cdot \nonumber \\
&\left[ 2a_q +r-R_q-4(r-R_q)\exp[(r-R_q)/a_q] \right. \nonumber \\
&\left. + (r-R_q-2a_q)\rme^{2(r-R_q)/a_q}) \right].
\eeqaa

Replacing in equation  (\ref{N fixed 2}) we get:

\beqa
\frac{\partial R_q}{\partial a_q} = \left(\frac{3}{4\pi}\frac{N_q}{\rho_{cq}} \right)^{1/3}
\left[ -\frac{2\pi^2 a_q}{3} \left(\frac{4\pi \rho_{cq}}{3N_q} \right)^{2/3} \right];
\eeqa

\beqa
\frac{\partial R_q}{\partial \rho_{cq}} = \left(\frac{3}{4\pi}\frac{N_q}{\rho_{cq}} \right)^{1/3}
\left[ -\frac{1}{3\rho_{cq}}-\frac{\pi^2 a_q^2}{9} \rho_{cq}^{-1/3}
\left(\frac{4\pi}{3N_q} \right)^{2/3} \right].
\eeqa
Finally, we can put all together in  equation  (\ref{eq zero}).

\section{Variational derivation above neutron drip}\label{appB}

For variational calculations above the neutron drip, the input quantities are the total baryonic density, denoted in this section by $\bar{\rho}_B$, and the proton fraction $Y_p$ in the WS cell. We have three extra variational variables to be determined, namely, the background densities $\rho_{bn},~\rho_{bp}$ and the WS radius $R_\mathrm{WS}$.
Denoting by $E$ the total ETF energy, the derivatives that remain to be calculated are:

\begin{equation}
\frac{\partial E}{\partial \rho_{bq}} =
4\pi \int_0^{R_\mathrm{WS}} \left( \left. \frac{\partial {\mathcal H}}{\partial \rho_{bq}}
\right|_{{R_q}=const} +\frac{\partial {\mathcal H}}{\partial R_q}
\frac{\partial R_q}{\partial \rho_{bq}} \right) r^2 dr;
\end{equation}

\begin{equation}
\frac{\partial}{\partial R_\mathrm{WS}} \left(\frac{E}{A} \right)= \frac{1}{A}\frac{\partial E}{\partial R_\mathrm{WS}}
- \frac{E}{A^2}\frac{\partial A}{\partial R_\mathrm{WS}},
\end{equation}
where:
\beqaa 
\frac{\partial E}{\partial R_\mathrm{WS}} =& 4\pi {\mathcal H}(R_\mathrm{WS})R_\mathrm{WS}^2 \nonumber \\
&+ 4\pi \int_0^{R_\mathrm{WS}} \left( \frac{\partial {\mathcal H}}{\partial R_n}
\frac{\partial R_n}{\partial R_\mathrm{WS}} + \frac{\partial {\mathcal H}}{\partial R_p}
\frac{\partial R_p}{\partial R_\mathrm{WS}} \right) r^2 dr;
\eeqaa
\begin{equation}
\frac{\partial A}{\partial R_\mathrm{WS}}= \bar{\rho}_B \, 4 \pi R_\mathrm{WS}^2;
\end{equation}
\begin{equation}
\frac{\partial R_q}{\partial \rho_{bq}}= \frac{\partial R_q}{\partial N_{C,q}}
\frac{\partial N_{C,q}}{\partial \rho_{bq}}= \frac{\partial R_q}{\partial N_{C,q}}
\left( -\frac{4\pi}{3} R_\mathrm{WS}^3 \right);
\end{equation}
\begin{equation}
\frac{\partial R_n}{\partial R_\mathrm{WS}}= \frac{\partial R_n}{\partial N_{C,n}}
(\bar{\rho}_B y_n-\rho_{bn})\, 4\pi R_\mathrm{WS}^2;
\end{equation}
\begin{equation}
\frac{\partial R_p}{\partial R_\mathrm{WS}}= \frac{\partial R_p}{\partial N_{C,p}}
(\bar{\rho}_B y_p-\rho_{bp})\, 4\pi R_\mathrm{WS}^2;
\end{equation}
\beqaa 
\frac{\partial R_q}{\partial N_{C,q}}=& \left(\frac{3}{4\pi}\frac{1}{\rho_{cq}} \right)^{1/3}
\left[ 1-\frac{\pi^2 a_q^2}{3} \left(\frac{4\pi \rho_{cq}}{3N_{C,q}} \right)^{2/3} \right]
\frac{N_{C,q}^{-2/3}}{3} \nonumber \\
&+\left(\frac{3}{4\pi}\frac{N_{C,q}}{\rho_{cq}} \right)^{1/3}
 \left[ \frac{2\pi^2 a_q^2}{9} \left(\frac{4\pi \rho_{cq}}{3} \right)^{2/3} N_{C,q}^{-5/3} \right],
\eeqaa 
where $y_n=1-y_p$ is the neutron fraction and $N_{C,q}=N_{q}-V_\mathrm{WS}\rho_{bq}$.
 
\section{Bulk Coulomb energy} \label{appC}

In this Appendix, we work out explicitly the expression of the bulk Coulomb energy 
equation (\ref{ex coulomb uniform 2}) in the presence of a proton density profile and a neutralizing homogeneous electron (and free proton)  background.  
We start from the direct and exchange Coulomb energy density equations (\ref{eq:exchange}) and (\ref{eq:couldens}):

\beqaa 
{\mathcal H}_\mathrm{Coul}=& \frac{e^2}{2} (\rho_p(r)-\rho_e)
\left[ \int_0^r \rho_p(r') \left( \frac{{r'}^2}{r}-r' \right) dr' +
\rho_e \frac{r^2}{6}\right]
\nonumber \\
& -\frac{3e^2}{16 \pi} \left(\frac{3}{\pi} \right)^{1/3}
(\rho_p^{4/3}(r) + \rho_e^{4/3}).
\eeqaa 

The bulk part is defined by introducing a sharp radius $R_{cl}$ and the associated bulk density $\rho_{cl,p}$ as:

\beqa
{\mathcal H}_\mathrm{Coul}=-\frac{e^2}{12} r^2 (\rho_{cl,p}- \rho_e)^2 -\frac{3e^2}{16 \pi}
\left(\frac{3}{\pi} \right)^{1/3}
(\rho_{cl,p}^{4/3} + \rho_e^{4/3}), \quad r \leq R_{cl};
\eeqa

\beqaa 
{\mathcal H}_\mathrm{Coul}= \frac{e^2}{2} (\rho_{bp}-\rho_e)&\left[ \rho_e \frac{r^2}{6} +\left( \frac{R_{cl}^3}{3r} -\frac{R_{cl}^2}{2}\right) \rho_{cl,p}
\right. \nonumber \\
& \left. -( r^2 -R_{cl}^2)\frac{\rho_{bp}}{6} \right], \quad r>R_{cl}.
\eeqaa 

The Coulomb energy results:

\beqa
E_\mathrm{Coul}= 4\pi \int_0^{R_{cl}} {\mathcal H}_\mathrm{Coul} r^2 dr
+ 4\pi \int_{R_{cl}}^{R_\mathrm{WS}} {\mathcal H}_\mathrm{Coul} r^2 dr;
\eeqa

\beqaa \label{ex coulomb uniform}
E_\mathrm{Coul}=& -\frac{4\pi e^2}{12} (\rho_{cl,p}-\rho_e)^2\frac{R_{cl}^5}{5}
-\frac{3e^2}{16 \pi} \left(\frac{3}{\pi} \right)^{1/3} \frac{4\pi}{3} \cdot \nonumber \\
& \left(\rho_{cl,p}^{4/3} R_{cl}^3 +\rho_{bp}^{4/3} (R_\mathrm{WS}^3 -
R_{cl}^3)  + \rho_e^{4/3} R_\mathrm{WS}^3\right) \nonumber \\
&-\frac{4\pi e^2}{12} (\rho_{bp}-\rho_e)^2\frac{R_\mathrm{WS}^5 -R_{cl}^5}{5}
\nonumber \\
&+ \frac{4\pi e^2}{12} (\rho_{bp}-\rho_e) \left[(R_{cl}^3 R_\mathrm{WS}^2
-R_{cl}^2 R_\mathrm{WS}^3) \rho_{cl,p} \right. \nonumber \\
&\left. + (R_{cl}^2 R_\mathrm{WS}^3 -R_{cl}^5) \frac{\rho_{bp}}{3}\right].
\eeqaa 

For $\rho_{bp}=0$, using
$Z=(4/3)\pi R_{cl}^3 \rho_{cl,p}=(4/3)\pi R_\mathrm{WS}^3 \rho_e$, we finally get:
\beqaa 
E_\mathrm{Coul}=& \frac{3}{5} \frac{e^2}{4\pi} \frac{Z^2}{R_{cl}}
\left( 1-\frac{3}{2} \frac{R_{cl}}{R_\mathrm{WS}}
+\frac{1}{2} \frac{R_{cl}^3}{R_\mathrm{WS}^3} \right) \nonumber \\
& -\frac{3e^2}{16 \pi} \left(\frac{3}{\pi} \right)^{1/3} Z
(\rho_{cl,p}^{1/3} + \rho_e^{1/3}).
\eeqaa

\section*{References}
\bibliography{bibliJPG_arxiv}
\bibliographystyle{h-elsevier}

\end{document}